\documentclass{jpp}
\usepackage{graphicx}
\usepackage{epstopdf, epsfig}
\def\ga{\mathrel{\mathchoice {\vcenter{\offinterlineskip\halign{\hfil
$\displaystyle##$\hfil\cr>\cr\sim\cr}}}
{\vcenter{\offinterlineskip\halign{\hfil$\textstyle##$\hfil\cr>\cr\sim\cr}}}
{\vcenter{\offinterlineskip\halign{\hfil$\scriptstyle##$\hfil\cr>\cr\sim\cr}}}
{\vcenter{\offinterlineskip\halign{\hfil$\scriptscriptstyle##$\hfil
\cr>\cr\sim\cr}}}}}
\def\la{\mathrel{\mathchoice {\vcenter{\offinterlineskip\halign{\hfil
$\displaystyle##$\hfil\cr<\cr\sim\cr}}}
{\vcenter{\offinterlineskip\halign{\hfil$\textstyle##$\hfil\cr<\cr\sim\cr}}}
{\vcenter{\offinterlineskip\halign{\hfil$\scriptstyle##$\hfil\cr<\cr\sim\cr}}}
{\vcenter{\offinterlineskip\halign{\hfil$\scriptscriptstyle##$\hfil
\cr<\cr\sim\cr}}}}}

\shorttitle{A  ``metric'' semi-Lagrangian Vlasov-Poisson solver}
\shortauthor{S. Colombi and C. Alard}

\title{A ``metric'' semi-Lagrangian Vlasov-Poisson solver}

\author{St\'ephane Colombi\aff{1,2}
  \corresp{\email{colombi@iap.fr}}
\and Christophe Alard\aff{1}}
\affiliation{\aff{1}Institut d'Astrophysique de Paris, CNRS UMR 7095 and UPMC, 98bis bd Arago, F-75014 Paris, France
\aff{2}Center for Gravitational Physics, Yukawa Institute for Theoretical Physics, Kyoto University, Kyoto 606-8502, Japan}
\begin{document}
\date{\today}
\maketitle
\begin{abstract}
We propose a new semi-Lagrangian Vlasov-Poisson solver. It employs elements of metric to follow locally the flow and its deformation, allowing one to find quickly and accurately the initial phase-space position ${\boldsymbol Q}({\boldsymbol P})$ of any test particle ${\boldsymbol P}$, by expanding at second order the geometry of the motion in the vicinity of the closest element. It is thus possible to reconstruct accurately the phase-space distribution function at any time $t$ and position ${\boldsymbol P}$ by proper interpolation of initial conditions, following Liouville theorem. When distorsion of the elements of metric becomes too large, it is necessary to create new initial conditions along with isotropic elements and repeat the procedure again until next resampling.  To speed up the process, interpolation of the phase-space distribution is performed at second order during the transport phase, while third order splines are used at the moments of remapping.  We  also show how to compute accurately the region of influence of each element of metric with the proper percolation scheme. The algorithm is tested here in the framework of one-dimensional gravitational dynamics but is implemented in such a way that it can be extended easily to four or six-dimensional phase-space. It can also be trivially generalised to plasmas. 
\end{abstract}
\section{Introduction}
The dynamics of Dark matter in the Universe and stars in galaxies is governed by Vlasov-Poisson equations
\begin{eqnarray}
\frac{\partial f}{\partial t}+{\boldsymbol v.\nabla_{\boldsymbol x}} f-\nabla_{\boldsymbol x} \phi\,. \nabla_{\boldsymbol v}f=0, \\
\Delta_{\boldsymbol x} \phi=4 \pi G \rho, \quad \rho({\boldsymbol x},t) \equiv \int f({\boldsymbol x},{\boldsymbol v},t)\, {\rm d}{\boldsymbol v},
\end{eqnarray}
where $f({\boldsymbol x}, {\boldsymbol v},t)$ is the phase-space distribution function at position ${\boldsymbol x}$, velocity ${\boldsymbol v}$ and time $t$, $\phi$ the gravitational potential, $G$ the gravitational constant and $\rho$ the projected density.  Usually, these equations are solved numerically with a $N$-body approach, where the phase-space distribution function is represented by an ensemble of Dirac functions interacting gravitationally with a softened potential \citep[see, e.g.,][and references therein]{Hockney1988,Bertschinger1998,Colombi2001,Dolag2008,Dehnen2011}. It is known that such a rough representation of the phase-space fluid is not free of biases nor systematic effects \citep[see, e.g.,][]{Aarseth1988,Kandrup1991,Boily2002,Binney2004,Melott2007,Joyce2009,Colombi2015}. With the computational power available today, an alternative approach is becoming possible, consisting in solving directly Vlasov dynamics in phase-space \citep[see, e.g.,][]{Yoshikawa2013}. In fact, this point of view is already commonly adopted in plasma physics \citep[see, e.g.,][and references therein]{Grandgirard2016}, but seldom used in the gravitational case. 

In this paper, we consider the case where the phase-space distribution is a smooth function of the vector 
\begin{equation}
{\boldsymbol P}\equiv({\boldsymbol x},{\boldsymbol v}). 
\end{equation}
This means that initial conditions have non zero initial velocity dispersion, which is relevant to the analysis of e.g. galaxies or relaxed dark matter halos. On the other hand, if the aim would be to follow accurately the dynamics of dark matter in the Cold Dark Matter paradigm, the initial velocity would be null. In this case, the problem is different since one has to describe the foldings of a $D$-dimensional hypersurface evolving in 2$D$ dimensional phase-space \citep[see, e.g.,][]{Hahn2016,Sousbie2016}.

In the warm case considered here, it is possible to represent $f({\boldsymbol P},t)$ on an Eulerian grid. Most of the numerical methods trying to resolve directly the dynamics of function $f({\boldsymbol P},t)$ have been developed in plasma physics and are of {\rm semi-Lagrangian} nature. They usually exploit directly Liouville theorem, namely that $f$ is conserved along characteristics. In practice, a test particle is associated to each grid site where the value $f_{\rm T}$ of the phase-space distribution function has to be evaluated. This test particle is followed backwards in time to find its position at previous time step. Then,  function $f$ from previous time step is interpolated at the root of the characteristic to obtain the searched value of $f_{\rm T}$, following Liouville theorem. This approach was pioneered by \citet{Cheng1976} in a split fashion and first applied to astrophysical systems by \citet{Fujiwara1981}, \citet{Nishida1981} and \citet{Watanabe1981}. In the classical implementation, interpolation of the phase-space distribution function is performed using a third order spline. 

There exist many subsequent improvements over the splitting algorithm of Cheng \& Knorr \citep[see, e.g.,][but this list is far from comprehensive]{Shoucri1978,sonnendrucker1999,Filbet2001,Besse2003,Crouseilles2010,Rossmanith2011,Qiu2011,Glucu2014}. The main problem common to all these algorithms is the unavoidable diffusion due to repeated resampling of the phase-space distribution function. Indeed, because of the finite grid resolution, repeated interpolation, as sophisticated it can be, induces information loss and augmentation of entropy. One way to fix this problem is to employ adaptive mesh refinement \citep[see, e.g.,][]{Gutnic2004,Besse2008,Besse2017}, i.e. to augment local resolution when the level of details asks for it. However, diffusion can also be significantly dampened if remapping of the phase-space distribution function is performed as seldom as possible. One way to achieve this is to represent $f({\boldsymbol P},t)$ with an ensemble of elements of which the shape is allowed to change with time according to the Lagrangian equations of motion \citep[see, e.g.,][]{Alard2005,CamposPinto2014,Larson2015}. In the ``cloudy'' method that we proposed in \citet[][]{Alard2005}, the phase-space distribution is sampled on an ensemble of overlapping round Gaussian functions of which the contours tranform into ellipsoids following the phase-space flow. When the ellipsoids become too elongated, the phase-space distribution function is resampled with new round clouds. Because resampling is seldom performed, say at fractions of dynamical times, diffusion effects are much less prominent, even though coarse graining is still performed at the resolution scale given by the inter-cloud center spacing. The problem with this remapping approach is the computational cost: to have a smooth and accurate representation of the phase-space distribution function, the clouds need to overlap significantly: the extra cost due to this overlap becomes prohibitive in high number of dimensions. 

However, one can realize in fact that it is not necessary to represent the phase-space distribution on a set of clouds: the only important piece of information is the metric intrinsically associated to each cloud of the previous approach, i.e. the information on how function $f({\boldsymbol P},t)$ is deformed by the motion. If one is indeed able to follow the properties of the motion up to some order in the neighbourhood of test particles, that we call metric elements, reconstructing the initial position ${\boldsymbol Q}({\boldsymbol P},t)$ of any point nearby is straightforward, and so is the calculation of  $f({\boldsymbol P},t)$ from interpolation of initial conditions following Liouville theorem. 

Our algorithm, {\tt Vlamet}, described in details in \S~\ref{sec:alg}, can thus be summarised as follows. It consists in sampling at all times the phase-space distribution function $f({\boldsymbol P},t)$ on a grid of fixed resolution  and following $f$ at second order in space and time using a predictor-corrector time integration scheme.  The calculation of $f({\boldsymbol P},t)$ exploits directly Liouville theorem, namely that $f$ is conserved along the characteristics, i.e. is equal to its initial value $f_{\rm ini}\equiv f({\boldsymbol Q},t_{\rm ini})$ at Lagrangian position ${\boldsymbol Q}({\boldsymbol P},t)$. To follow these characteristics accurately, we define a set of elements of metric that are sparse sampling the grid. These elements of metric follow the Lagrangian equations of motion and carry during time local information about the deformation of phase-space up to second order,  to account for the curvature of phase-space structures building up during the course of the dynamics.  To resample $f$ at time $t$ and at a given point ${\boldsymbol P}$ of phase-space, we consider the closest elements of metric in Lagrangian space and reconstruct the initial position ${\boldsymbol Q}$ of a test particle coinciding with ${\boldsymbol P}$ using a second order expansion of the geometry of the flow around the metric elements. Once the initial position ${\boldsymbol Q}$ is found, we just need to interpolate $f_{\rm ini}$ at position ${\boldsymbol Q}$ and attribute it to $f({\boldsymbol P},t)$. 

\begin{figure}
\begin{center}
\includegraphics[width=0.7\linewidth]{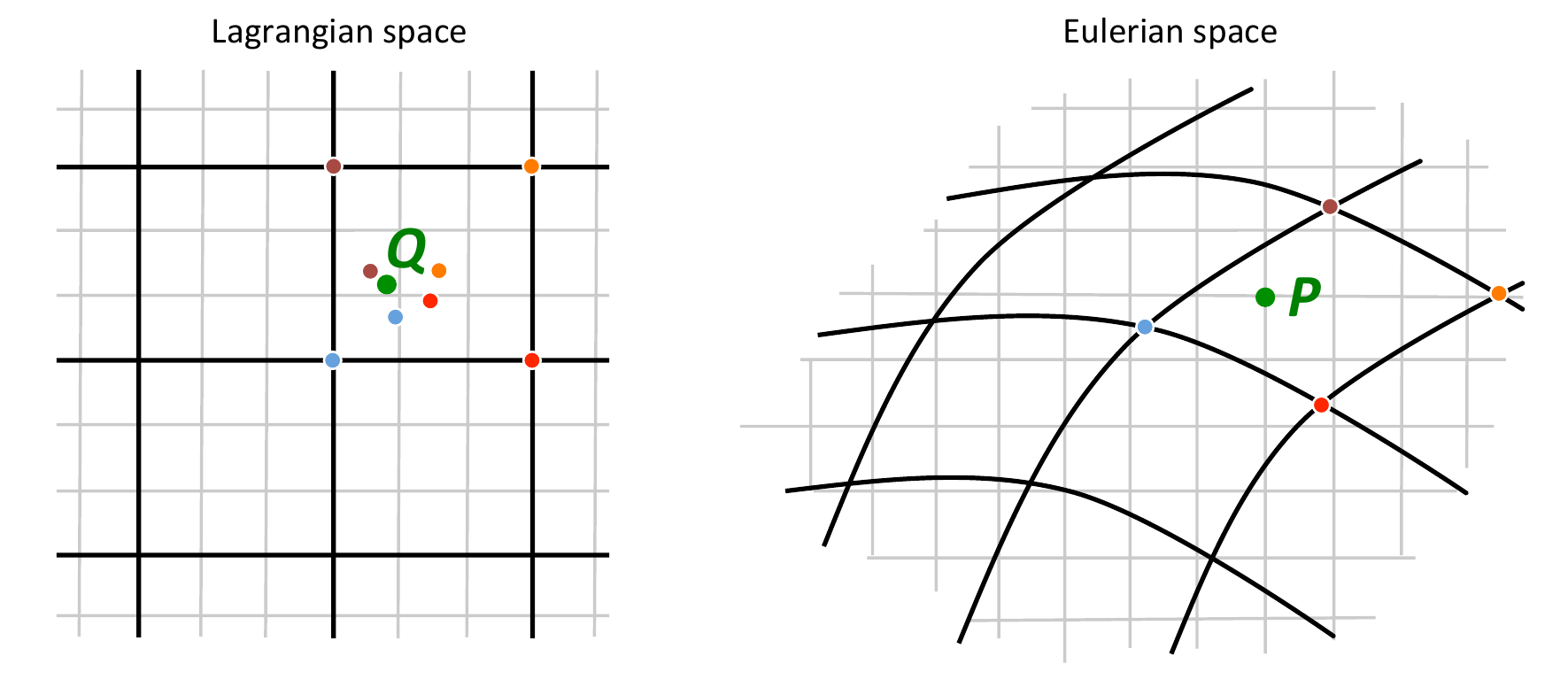}
\end{center}
\caption[]{Schematic representation of the algorithm. At all times we would like to follow the phase-space distribution function $f({\boldsymbol P},t)$ sampled on a fine Eulerian grid, represented here in grey. To achieve this, we define local elements of metric initially sampling a sparser grid, as represented in black on left panel. These elements of metric follow the motion and its distorsions, as illustrated by the deformation of the black grid on right panel. To reconstruct $f$ at all times on the Eulerian grid, e.g. the green point on right panel, one traces back in time the trajectory of a test particle associated to the green point. To compute the position of this particle during the transport phase, we just use the closest element of metric in Lagrangian space, here the blue point on left panel. Once ${\boldsymbol Q}$ is calculated, we apply Liouville theorem i.e., $f({\boldsymbol P},t)=f_{\rm ini}\equiv f({\boldsymbol Q},t_{\rm ini})$ where $t_{\rm ini}$ is the initial time, computing $f({\boldsymbol Q},t_{\rm ini})$ with a second order interpolation over the Lagrangian grid.  Of course, at some point, $t=t_{\rm resample}$, distortion of phase-space structures is too large to be only locally represented at second order, so one has to start again the process by considering $f({\boldsymbol P},t_{\rm resample})$ as new initial conditions. At time $t_{\rm resample}$, we proceed slightly differently to compute $f({\boldsymbol P},t)$ more accurately. Indeed, because our local representation of phase-space is only valid up to second order, the reconstructed initial position proposed by each element of metric is slightly different, as illustrated by left panel. To avoid an unsmooth representation of the reconstruction of the initial position ${\boldsymbol Q}$, we perform at $t=t_{\rm resample}$ a more accurate interpolation of the initial position, here between the purple, the orange, the red and the blue points and a more accurate interpolation using third order splines on the Lagrangian grid to compute $f({\boldsymbol Q},t_{\rm ini})$.}
\label{fig:recons}
\end{figure}

At some time $t_{\rm resample}$, the distorsion of phase-space is too strong to be simply represented at second order: $f({\boldsymbol P},t_{\rm resample})$ is considered as a new initial condition and new isotropic elements of metric are created. Between initial time and $t_{\rm resample}$, we use a cheap, second order interpolation to reconstruct $f({\boldsymbol P},t)$, while resampling at $t=t_{\rm resample}$ is more accurate, using third order splines.  Indeed, although our algorithm is meant to be globally precise only at second order, a third order description of the phase-space distribution function is essential at the moment of resampling to be able to conserve quantities such total energy and total mass in the long term.  Furthermore, to preserve smoothness of the resampling procedure, we perform some interpolation in Lagrangian space between the initial positions ${\boldsymbol Q}$ proposed by each element of metric close to the point of interest. Indeed, different neighbouring metric elements can provide different answers.  The procedure is illustrated by Fig.~\ref{fig:recons}. 

In \S~\ref{sec:numexp}, we thoroughly test the performances of our algorithm, by checking how a stationary solution is preserved and by simulating systems with Gaussian initial condition as well as random sets of halos.  To demonstrate the potential improvements brought by our code {\tt Vlamet} compared to traditional semi-Lagrangian approaches, we compare our algorithm to the traditional splitting method of \citet{Cheng1976} with third order spline interpolation. Results are also tested against simulations performed with the entropy conserving waterbag code of \citet[][]{Colombi2014}. The waterbag scheme, extremely accurate but very costly, is meant to provide a supposedly ``exact'' solution. It is mainly used to test the convergence of {\tt Vlamet} and of the splitting algorithm for the Gaussian initial condition. From the pure mathematical side, which is not really approached in this article, it is worth mentioning the recent work of \citet[][]{CamposPinto2016}, who discuss some convergence properties of a simplified version of our algorithm. 
\section{Algorithm}
\label{sec:alg}
Now we describe the main steps of the algorithm, namely, (i) our second order representation of the metric (\S~\ref{sec:2nd}), (ii) our interpolation schemes to reconstruct $f$ (\S~\ref{sec:int}), (iii) our percolation algorithm to find the nearest element of metric to an Eulerian pixel ${\boldsymbol P}$ and the reconstruction of its Lagrangian counterpart ${\boldsymbol Q}$ using an interpolation between the alternatives proposed by the neighbouring elements of metric (\S~\ref{sec:recQ}), (iv) the equations of motion as well as their numerical implementation  (\S~\ref{sec:moti}) and finally, (v) the calculation of the force, its first and second derivatives, required to follow the second order representation of the local metric during motion (\S~\ref{sec:forc}). 

While the actual implementation of the algorithm is performed in 2D phase-space, generalisation to higher number of dimensions will be discussed in details. Indeed, our algorithm is meant to improve other standard semi-Lagrangian solvers in terms of diffusion while being of comparable cost, and is designed in such a way that it can be easily generalised to 4- and 6-dimensional phase-space.  

\subsection{Second order representation of the metric}
\label{sec:2nd}
The Lagrangian position of each fluid element in phase space is defined by
\begin{equation}
{\boldsymbol Q} \equiv (q_x,q_v),
\end{equation}
where $q_x$ and $q_v$ are its initial position and velocity, respectively. Here we consider 2 dimensional phase-space, so these quantities are scalar, but in general, they are vectors. In what follows, unless specified otherwise, all the calculations apply to the general case: one just has to replace $q_x$ and $q_v$ with ${\boldsymbol q}_x$ and ${\boldsymbol q}_v$. 
The Eulerian position at time $t$ of an element of fluid in phase-space is given by 
\begin{equation}
{\boldsymbol P}({\boldsymbol Q},t) \equiv [x(q_x,q_v,t),v(q_x,q_v,t)],
\end{equation}
where $x(q_x,q_v,t)$ and $v(q_x,q_v,t)=\partial x/\partial t$ are its position and velocity at time $t$. 
Initially
\begin{equation}
{\boldsymbol P}({\boldsymbol Q},t=0)={\boldsymbol Q}.
\end{equation}
In our second order approach, we need the deformation tensor,
\begin{equation}
{{{\boldsymbol T}}}({\boldsymbol Q},t) \equiv \frac{\partial {\boldsymbol P}}{\partial {\boldsymbol Q}},
\end{equation}
as well as the Hessian of the characteristics,
\begin{equation}
{{{{\boldsymbol H}}}}({\boldsymbol Q},t) \equiv \frac{\partial^2 {\boldsymbol P}}{\partial{\boldsymbol Q}^2}.
\end{equation} 
With the knowledge at all times of the position ${\boldsymbol P}_{\rm m}$, the deformation tensor ${\boldsymbol T}_{\rm m}$ and the Hessian ${\boldsymbol H}_{\rm m}$ for each element of metric, we can predict, at second order, the position of an element of fluid initially sufficiently close to an element of metric:
\begin{equation}
{\boldsymbol P} \simeq{\boldsymbol P}_{\rm m}+{\boldsymbol T}_{\rm m} ({\boldsymbol Q}-{\boldsymbol Q}_{\rm m})+\frac{1}{2} ({\boldsymbol Q}-{\boldsymbol Q}_{\rm m})^{\rm T} {\boldsymbol H}_{\rm m} ({\boldsymbol Q}-{\boldsymbol Q}_{\rm m}).
\end{equation}
Of course the series expansion could be developed further, but we restrict here to an algorithm at second order in space and time. Indeed, going beyond second order would be very costly in higher number of dimensions, particularly 6-dimensional phase-space. The interesting bit, as we shall see later, is the inverse relation between Eulerian position and Lagrangian position, i.e., at second order
\begin{eqnarray}
{\boldsymbol Q} &\simeq & {\boldsymbol Q}_{\rm m}+{\boldsymbol T}_{\rm m}^{-1}({\boldsymbol P}-{\boldsymbol P}_{\rm m}) -\frac{1}{2}{\boldsymbol T}_{\rm m}^{-1} ({\boldsymbol P}-{\boldsymbol P}_{\rm m})^{\rm T}[ {\boldsymbol T}_{\rm m}^{-1}]^{\rm T}{\boldsymbol H}_{\rm m} {\boldsymbol T}_{\rm m}^{-1} ({\boldsymbol P}-{\boldsymbol P}_{\rm m}).\label{eq:QofP}
\end{eqnarray}

In the two dimensional phase-space case considered here, the deformation tensor ${\boldsymbol T}$ is a 4 elements matrix {\em with no special symmetry} except that according to the Hamiltonian nature of the system  its determinant should be conserved with time and equal to the initial value:
\begin{equation}
|{{{\boldsymbol T}}}|=1,
\end{equation}
at all times. Indeed 
\begin{equation}
{{{\boldsymbol T}}}({\boldsymbol Q},t=0)={\boldsymbol I},
\end{equation}
where ${\boldsymbol I}$ is the identity matrix. 
Note that this property is general, as volume is conserved in phase-space, it is not specific to the one dimensional dynamical case. We have,
\begin{equation}
J\equiv |{{\boldsymbol T}}|=\frac{\partial x}{\partial q_x}\frac{\partial v}{\partial q_v}-\frac{\partial x}{\partial q_v}\frac{\partial v}{\partial q_x}=1, \label{eq:detJ}
\end{equation}
hence ${\boldsymbol T}$ has only 3 independent elements. In practice we shall not impose equation (\ref{eq:detJ}) but checking to which extent it is verified can be used as a test of the good numerical behaviour of the code, or equivalently, as a test of its symplecticity.

In higher number of dimensions, phase-space volume conservation is not the only constraint. In fact it is a consequence of the symplectic nature of the system, which hence conserves Poincar\'e invariants, in particular symplectic forms of the kind
\begin{equation}
{\cal I}=\frac{1}{2}[{{\rm d} {\boldsymbol P}_1}]^{\rm T}\, {\boldsymbol S}\, {\rm d}{\boldsymbol P}_2,
\end{equation}
where 
\begin{equation}
{\boldsymbol S} \equiv \left( \begin{array}{cc} {\boldsymbol 0} & -{\boldsymbol I} \\ {\boldsymbol I} & {\boldsymbol 0} \end{array}\right)
\end{equation}
is the symplectic matrix, while ${\rm d} {\boldsymbol P}_1$ and ${\rm d} {\boldsymbol P}_2$ correspond to 2 sides of an elementary triangle composed of three particles following the equations of motion in phase-space. In particular,
\begin{equation}
{\rm d} {\boldsymbol P}_j = \frac{\partial {\boldsymbol P}}{\partial {\boldsymbol Q}}\, {\rm d} {\boldsymbol Q}_j.
\end{equation}
So the generalisation of equation (\ref{eq:detJ}) to higher number of dimensions provides the following constraints:
\begin{eqnarray}
\left[\frac{\partial {\boldsymbol v}}{\partial {\boldsymbol q}_x}\right]^{\rm T} \frac{\partial {\boldsymbol x}}{\partial {\boldsymbol q}_x} 
-  \left[\frac{\partial {\boldsymbol x}}{\partial {\boldsymbol q}_x} \right]^{\rm T}\frac{\partial {\boldsymbol v}}{\partial {\boldsymbol q}_x} = 0, \label{eq:poinca1} \\
\left[\frac{\partial {\boldsymbol v}}{\partial {\boldsymbol q}_v}\right]^{\rm T} \frac{\partial {\boldsymbol x}}{\partial {\boldsymbol q}_v} 
-  \left[\frac{\partial {\boldsymbol x}}{\partial {\boldsymbol q}_v}\right]^{\rm T} \frac{\partial {\boldsymbol v}}{\partial {\boldsymbol q}_v} = 0, \label{eq:poinca2} \\
 \left[\frac{\partial {\boldsymbol x}}{\partial {\boldsymbol q}_x} \right]^{\rm T} \frac{\partial {\boldsymbol v}}{\partial {\boldsymbol q}_v} - \left[\frac{\partial {\boldsymbol v}}{\partial {\boldsymbol q}_x}\right]^{\rm T} \frac{\partial {\boldsymbol x}}{\partial {\boldsymbol q}_v} = {\boldsymbol I}, \label{eq:poinca3}
\end{eqnarray}
which can be again used to double check, a posteriori, the good behaviour of the code in the general case. 

In the 2-dimensional phase-space case considered here, the Hessian ${{{\boldsymbol H}}}$ is a $2\times 2 \times 2$ tensor in which only 6 terms are independant if one just takes into account the symmetries in second derivatives, e.g. $\partial^2 x/(\partial q_x \partial q_v)=\partial^2x/(\partial q_v \partial q_x)$. In 4- and 6- dimensional phase-space, the same symmetries imply that among its $4\times 4 \times 4=64$ and $6 \times 6 \times 6=$ 216 elements, the Hessian includes 40 and 126 independant terms, respectively.

While exploiting the Hamiltonian nature of the system would allow us to carry less information, we chose for each element of metric to store the phase-space coordinates, the deformation tensor and the Hessian. The calculations above show that this ends up in carrying 2+4+6=12, 4+16+40=60 and 6+36+126=168 numbers for each element of metric in 2-, 4- and 6-dimensional phase-space, respectively. This might look a large number, particularly in 4- and 6-dimensional phase-space, but one has to keep in mind the fact that elements of metric are much sparser than the grid used to compute the phase-space distribution function. For instance, a factor 3 in each direction of the spacing between elements of metric compared to the grid cell size implies that there is one element of metric for $3^6=729$ grid sites. Another issue is the cost of applying the inverse mapping (\ref{eq:QofP}), but it can be reduced considerably by precomputing for each element of metric the matrixes ${\boldsymbol T}_{\rm m}^{-1}$ and $[{\boldsymbol T}_{\rm m}^{-1}]^{\rm T} {\boldsymbol H}_{\rm m} {\boldsymbol T}_{\rm m}^{-1}$ prior to any interpolation of the phase-space distribution function using the inverse mapping. 

\subsection{Interpolation schemes}
\label{sec:int}
We use two interpolation schemes. Between two coarse steps, a period of time during which the phase-space distribution function is transported along motion using the same elements of metric at each time step, we use a simple interpolation of $f$ based on local fitting of a quadratic hypersurface. Then, during remapping, i.e. when new elements of metric are set, $f$ is interpolated with a third order spline in order to have a more accurate reconstruction of phase-space. The advantage of using second order interpolation during the transport phase is that generalisation to 6-dimensional phase-space is relatively cheap, since only 28 terms contribute in this case to such an interpolation. The loss of accuracy due to the relative roughness of the reconstruction, which is not even continuous, has little consequence on the dynamics because most of the information is recovered during the resampling phase. This will be tested and demonstrated in \S~\ref{sec:numexp}. On the other hand, because our method does not allow for splitting, $4^6=4096$ terms contribute to each grid site when resampling $f$ in 6 dimensions with the third order spline  interpolation, if one assumes that this latter takes about 4 operation per grid site in one dimension.  This rather costly step is thus roughly  $4096/24 \simeq 171$  times more expensive than what is expected when performing a full time step in the standard splitting algorithm.\footnote{{ This calculation assumes that third order spline interpolation is used in the splitting algorithm. In a split fashion, this interpolation is performed individually on each axis, which corresponds, in 6 dimension, to $6\times 4=24$ operations per grid site for a full time step, or $9 \times 4=36$ operations if synchronisation between velocities and positions is performed. }} 
It is however performed rarely, hence making the algorithm of  potentially  comparable cost to the standard semi-Lagrangian approach, with the advantage of being, in principle, less diffusive.

Now we provide the details for the second order interpolation we use, followed by the standard third order spline interpolation.
\subsubsection{Second order interpolation}
The general procedure for second order interpolation consists in understanding that a second order hypersurface requires respectively 6, 15 and 28 points in 2D, 4D and 6D phase-space. What we have access to are the values $F({\boldsymbol G})$ of the phase-space distribution function on a grid ${\boldsymbol G}$, e.g., in one dimension ${\boldsymbol G}=(i,j)$ where $i$ and $j$ are integers, corresponding to the following Lagrangian positions in phase space,
\begin{equation}
{\boldsymbol Q}_{\rm G}=( i\Delta x+q_{x,{\rm mid}}, j \Delta v+q_{v,{\rm mid}}),
\label{eq:gridgrid}
\end{equation}
where $\Delta x$ and $\Delta v$ are the steps chosen for the grid (Eulerian or Lagrangian) that remain, in the current version of the code, fixed during runtime and $(q_{x,{\rm mid}},q_{v,{\rm mid}})$ corresponds to the middle of the Lagrangian computational domain. The bounds of the computational domain are then related to maximum values $n_x$ and $n_v$ of $|i|$ and $|j|$. 

\begin{figure}
\begin{center}
\includegraphics[width=0.6\linewidth]{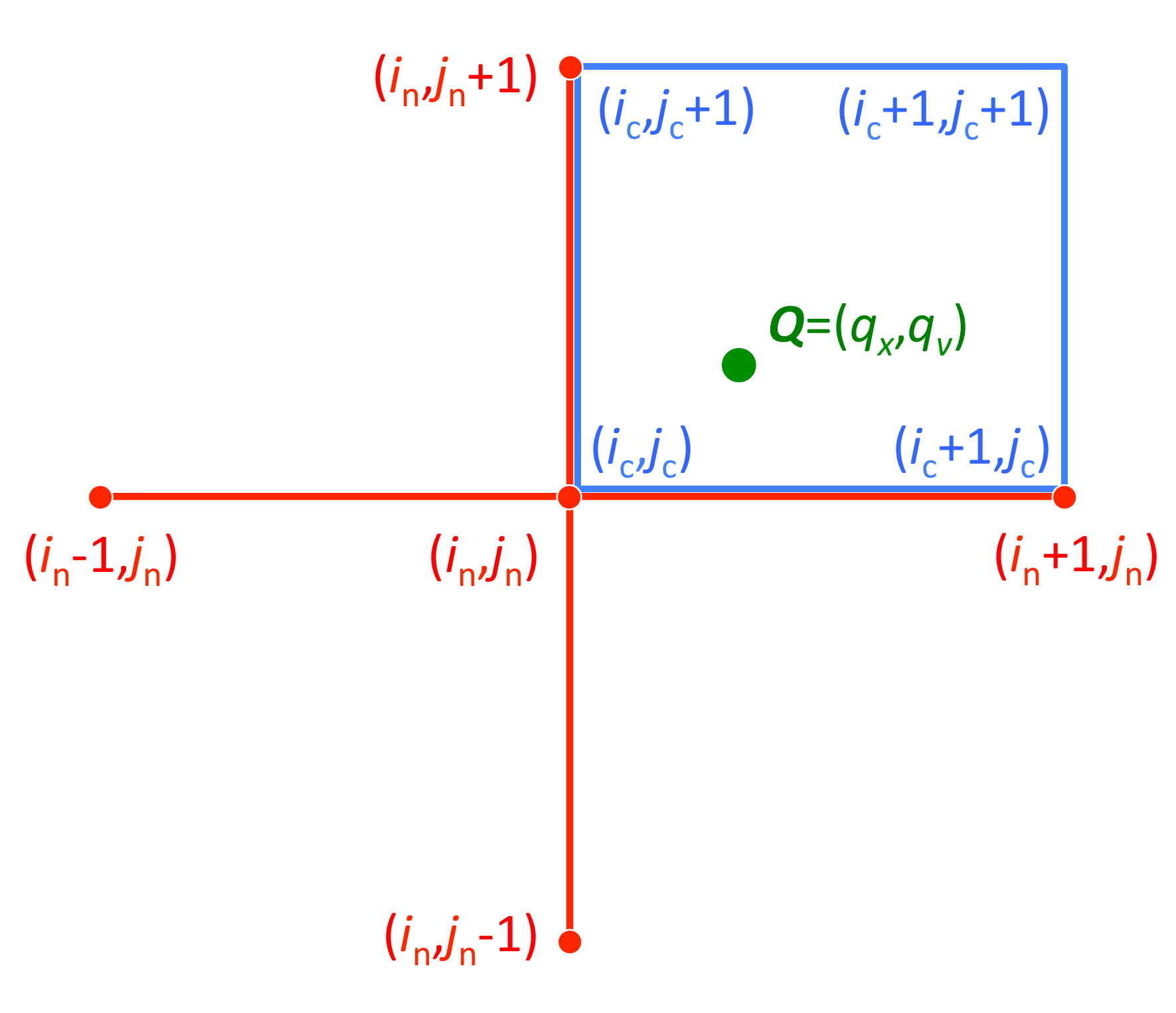}
\end{center}
\caption[]{Procedure employed to select the points of the grid used to perform second order interpolation of the phase-space distribution function at position ${\boldsymbol Q}$ in $2D$-dimensional phase-space. One first finds the nearest grid point ${\boldsymbol G}_{\rm n}$  to ${\boldsymbol Q}$, ${\boldsymbol G}_{\rm n}=(i_{\rm n},j_{\rm n})$ for $D=1$, and select all the segments of the grid composed of 3 grid points and centered on ${\boldsymbol G}_{\rm n}$: one obtains the red cross for $D=1$. Then one selects, for each plane passing through a pair of such segments the four elements of the grid so that the square composed of these grid elements contains the projection of ${\boldsymbol Q}$: one of them, $(i_{\rm c}+1,j_{\rm c}+1)$ on the figure, provides one additional point for the interpolation.}
\label{fig:schematicsinter}
\end{figure}

This is the way we proceed in 2D phase-space: given a point ${\boldsymbol Q}=(q_x,q_v)$ in phase space,  we first find the nearest grid point $(i_{\rm n},j_{\rm n})$ to it, then consider the cross composed of $(i_{\rm n}+1,j_{\rm n})$, $(i_{\rm n},j_{\rm n}+1)$, $(i_{\rm n}-1,j_{\rm n})$, $(i_{\rm n},j_{\rm n}-1)$ and $(i_{\rm n},j_{\rm n})$ on which the interpolation must coincide with $F$. The last point $(i,j)$ needed is taken to be the one for which a square composed of 3 points of the cross and $(i,j)$ contains $(q_x,q_v)$ (see Fig.~\ref{fig:schematicsinter}). With this choice of the points of the grid for the interpolation, we have, at second order, 
\begin{eqnarray}
f(q_x,q_v) &=& F(i_{\rm n},j_{\rm n})  (1-{\rm d}q_{x}^2-{\rm d}q_{v}^2 ) \nonumber \\
& &+ \frac{1}{2} F(i_{\rm n}+1,j_{\rm n}) {\rm d}q_x (1+{\rm d}q_x) + \frac{1}{2} F(i_{\rm n}-1,j_{\rm n}) {\rm d}q_x (1-{\rm d}q_x) \nonumber \\
& &+ \frac{1}{2} F(i_{\rm n},j_{\rm n}+1) {\rm d}q_v (1+{\rm d}q_v) + \frac{1}{2} F(i_{\rm n}-1,j_{\rm n}) {\rm d}q_v (1-{\rm d}q_v) \nonumber \\
& &+ [ F(i_{\rm c}+1,j_{\rm c}+1)+F(i_{\rm c},j_{\rm c}) -F(i_{\rm c}+1,j_{\rm c})-F(i_{\rm c},j_{\rm c}+1)] {\rm d}q_x {\rm d}q_v,\nonumber \\
\label{eq:int1}
\end{eqnarray}
with
\begin{eqnarray}
(i_{\rm n},j_{\rm n}) &=&(\lfloor(q_x-q_{x,{\rm mid}})/\Delta x\rceil,\lfloor (q_v-q_{v,{\rm mid}})/\Delta v\rceil), \\
(i_{\rm c},j_{\rm c}) &=& (\lfloor (q_x-q_{x,{\rm mid}})/\Delta x\rfloor, \lfloor (q_v-q_{v,{\rm mid}})/\Delta v \rfloor),\\
{\rm d}q_x &=& (q_x-q_{x,{\rm mid}})/\Delta x-i_{\rm n}, \\
{\rm d}q_v &=& (q_v-q_{v,{\rm mid}})/\Delta v-j_{\rm n},
\end{eqnarray}
where $\lfloor.\rceil$ and $\lfloor . \rfloor$ are respectively the nearest integer and the integer part functions.  Note obviously that when defining the edges of the computing domain, one makes sure that the interpolation is defined everywhere. In the present implementation, $f$ is supposed to be different from zero on a compact domain in phase-space. To avoid artificial enlarging of the computing domain during runtime due to diffusion and aliasing, the computing domain is restricted to a rectangular region containing all the points where
\begin{equation}
|f| > f_{\rm th}, \label{eq:constrf}
\end{equation}
with $f_{\rm th}$ a small number, typically $f_{\rm th}=10^{-6}$. Note the absolute value in equation (\ref{eq:constrf}), because our algorithm does not enforce positivity of $f$ and better total mass conservation is obtained by keeping the regions where $f < 0$. 

The generalisation of this second order interpolation to higher number of dimensions is straightforward: the most difficult is to chose the grid points through which the hypersurface must pass. To achieve this, similarly as in 2 dimensions, we first find the nearest grid point ${\boldsymbol G}_{\rm n}$ to ${\boldsymbol Q}$. Then take the collection of all segments centered on  ${\boldsymbol G}_{\rm n}$ and composed of 3 points of the grid (the cross defined above in 2 dimensions): this provides us in total 9 and 13 points respectively in 4 and 6 dimensions. Then, project ${\boldsymbol Q}$ on the planes defined by each pair of such segments and find the corresponding grid site in this plane such that the square composed of this grid site plus 3 contained in the 2 segments contains the projection of ${\boldsymbol Q}$ and add it to our list of points defining the second order hypersurface. There are 6 and 15 such points in 4 and 6 dimensions, respectively, and we thus end up using this procedure with respectively 15 and 28 points as needed to interpolate our hypersurface, with straightforward generalisation of equation (\ref{eq:int1}) to perform the interpolation. 

\subsubsection{Standard third order spline}
The third order interpolation we use employs standard cubic B-splines as very clearly described in e.g. \citet[][]{Crouseilles2009}, so we do not provide the details here, except the edge conditions, which are
\begin{eqnarray}
f(q_x,q_v) &=& 0, \quad \frac{\partial f}{\partial q_x}=0, \quad |q_x|=q_{x,{\rm mid}}+n_x \Delta x,\\
f(q_x,q_v) &=& 0, \quad \frac{\partial f}{\partial q_v}=0, \quad |q_v|=q_{v,{\rm mid}}+n_v \Delta v,
\end{eqnarray}
and this can be trivially generalised to 4- or 6-dimensional phase-space. 

\subsection{Tracing back the characteristics in Lagrangian space: percolation algorithm}
\label{sec:recQ}
To determine the value of $f({\boldsymbol P},t)$, we have to trace the trajectory of a test particle coinciding with ${\boldsymbol P}$ back to its Lagrangian, initial position ${\boldsymbol Q}$ (the position at time of remapping, in general). To do this, we need to find the closest element of metric to this test particle and use equation (\ref{eq:QofP}) to compute ${\boldsymbol Q}$. However, the closest element of metric, from the dynamical point of view, has to be found in Lagrangian space, which complicates the procedure. Furthermore, two close test particles might end up associated to different elements of metric that will propose slightly different mappings, implying that function ${\boldsymbol Q}({\boldsymbol P})$ is discontinuous if no supplementary treatment is performed. While this is not a big deal during the transport phase, it can have dramatic consequences on the long term during the remapping step which requires high level of precision when interpolating the phase-space distribution function with third order splines. Hence,  each time a full remapping of $f$ is performed, one needs to find not one but several elements of metric in the neighbourhood of a test particle to be able to interpolate between the proposed Lagrangian positions provided by each of these metric elements so that function ${\boldsymbol Q}({\boldsymbol P})$ stays smooth: this is the most complex part of the algorithm, because enforcing full continuity of the interpolation is actually not trivial since it has to be performed in Lagrangian space. The percolation algorithm however allows us to define in an unambiguous way regions of influence of each element of metric which in turn will allow us to make function ${\boldsymbol Q}({\boldsymbol P})$ smooth. 

We assume that elements of metric are initially set on a Lagrangian grid covering the computing domain, ${\boldsymbol M}=(i_{\rm m},j_{\rm m})$, corresponding to Lagrangian positions 
\begin{equation}
{\boldsymbol Q}_{\rm m}(i_{\rm m},j_{\rm m})=(\Delta x_{\rm m}\, i_{\rm m}+q_{x,{\rm mid}},\Delta v_{\rm m}\, j_{\rm m}+q_{v,{\rm mid}}),
\label{eq:gridcloud}
\end{equation}
with
\begin{equation}
\Delta x_{\rm m} \ga \Delta x,\quad \Delta v_{\rm m} \ga \Delta v.
\end{equation}
The actual computing domain changes subsequently with time. To estimate it at each time step we use the positions of the metric elements and we define a rectangle with lower-left and upper right corners respectively given by $[{\rm min}_{i_{\rm m},j_{\rm m}}(x_{\rm m}), {\rm min}_{i_{\rm m},j_{\rm m}}(v_{\rm m})]$ and $[{\rm max}_{i_{\rm m},j_{\rm m}}(x_{\rm m}), {\rm max}_{i_{\rm m},j_{\rm m}}(v_{\rm m})]$. This means that the element of metric coverage should be wide enough to avoid missing regions where $f$ should be non equal to zero. We therefore allow for an extra layer of metric elements on all the sides of the Lagrangian computing domain. Note thus that the calculation of the computing domain is far from being optimal, which can have dramatic consequences for the cost of the algorithm in 6 dimensions. In order to fix this problem, one would need to define a more complex shape dependent computational domain by e.g. constructing a k-d tree, i.e. a spatial decomposition of phase-space with nested cubes. Such a k-d tree could also constitute the basis for an adaptive mesh refinement algorithm following in an optimal way not only the computational domain, but also all the details of the phase-space distribution function when needed. However such a sophistication is not needed in our 2-dimensional implementation so we stick to the rectangular computing domain for now.

\subsubsection{Percolation}
\label{sec:perco}
To compute the region of influence of each element of metric $(i_{\rm m},j_{\rm m})$ with coordinates $(x_{\rm m},v_{\rm m})$, we percolate around it on the Eulerian computational grid starting from its nearest grid point 
\begin{equation}
(i,j)=\left(\lfloor (x_{\rm m}-x_{\rm mid})/\Delta x\rceil, \lfloor (v_{\rm m}-v_{\rm mid})/\Delta v\rceil \right),
\end{equation}
with $(x_{\rm mid},v_{\rm mid})$ being the time dependent position of the middle of the Eulerian computational domain. The initial Lagrangian position ${\boldsymbol Q}(i,j,i_{\rm m},j_{\rm m})$ of this point is estimated with equation (\ref{eq:QofP}) using the element of metric we started from as a reference. Note that another neighbouring element of metric $(i_{\rm m}',j_{\rm m}')$ would propose a slightly different value for the initial position of the test particle associated to Eulerian grid site $(i,j)$, hence the dependence on $(i_{\rm m},j_{\rm m})$ of ${\boldsymbol Q}$. 
Then one defines the function
\begin{eqnarray}
\delta {\boldsymbol Q}(i,j,i_{\rm m},j_{\rm m})  & \equiv & [\delta q_x(i,j,i_{\rm m},j_{\rm m}),\delta q_v (i,j,i_{\rm m},j_{\rm m})] \\
&\equiv & {\boldsymbol Q}(i,j,i_{\rm m},j_{\rm m}) -{\boldsymbol Q}_{\rm m}(i_{\rm m},j_{\rm m}), 
\end{eqnarray}
where ${\boldsymbol Q}_{\rm m}(i_{\rm m},j_{\rm m})$ is the Lagrangian position of the element of metric, 
to find out which neighbours $(i_{\rm m}',j_{\rm m}')$ of the element of metric are susceptible to be closer to the true initial ${\boldsymbol Q}(i,j)$ in Lagrangian space. Because the calculation of ${\boldsymbol Q}(i,j)$ is not exact we need to introduce the concept of uncertainty. Considering all the neighbours of the current element of metric under consideration
\begin{equation}
(i_{\rm m}',j_{\rm m}'):  \left\{ \begin{array}{ll} i_{\rm m}'\in [i_{\rm m}-1,i_{\rm m}+1], \\
 j_{\rm m}'\in [j_{\rm m}-1,j_{\rm m}+1], \\
(i_{\rm m}',j_{\rm m}') \neq (i_{\rm m},j_{\rm m}), \end{array} \right.
\end{equation}
we are able to chose the one which proposes the smallest value of 
\begin{equation}
d[\delta {\boldsymbol Q}(i,j,i_{\rm m},j_{\rm m})] \equiv \sqrt{(\delta q_x/\Delta x)^2 + (\delta q_v/\Delta v)^2},
\label{eq:distdef}
\end{equation}
where the dependence on $(i,j,i_{\rm m},j_{\rm m})$ of $\delta q_x$ and $\delta q_v$ has been made implicit.
If there is a candidate element of metric $(i_{\rm m}',j_{\rm m}')$ closer than $(i_{\rm m},j_{\rm m})$ to the Eulerian grid element $(i,j)$ using this Lagrangian distance, the percolation process is stopped. Otherwise, one continues (bond) percolation i.e. proceeds with neighbouring grid sites $(i+1,j)$, $(i-1,j)$, $(i,j+1)$, $(i,j-1)$ unless they have been already processed. 

Note that the percolation procedure just defined above is not always free of aliasing: it can be successful only if the separation between elements of metric is sufficiently large compared to the grid element size and if their zone of influence is not too distorted. Testing the validity of the percolation process is relatively simple, by checking if some points inside the computing domain have been left unexplored.  If it is the case, percolation is performed again locally until all the unexplored points are processed, which induces a small additional cost in the algorithm. 

\begin{figure}
\begin{center}
\includegraphics[width=0.6\linewidth]{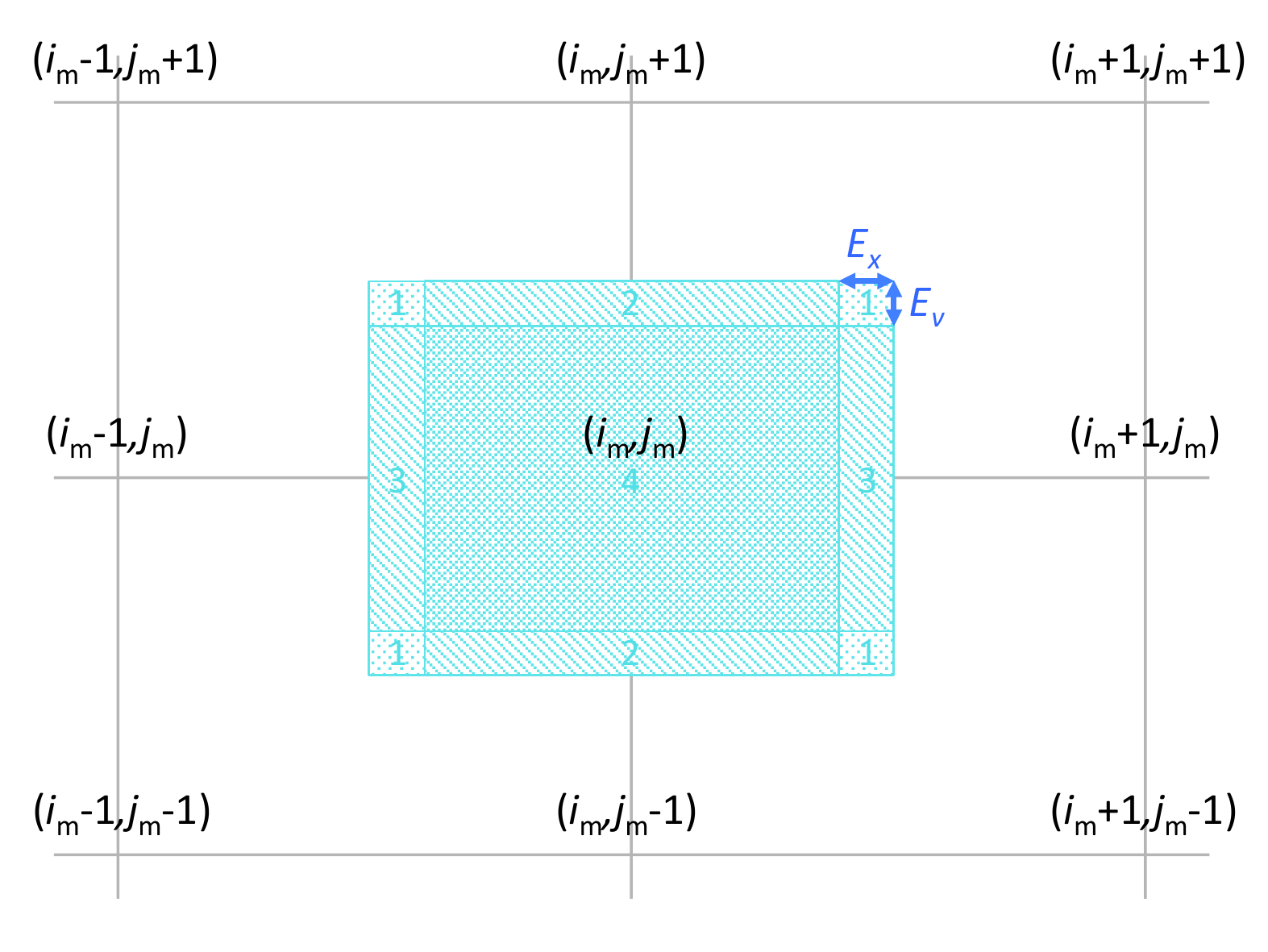}
\end{center}
\caption[]{Scheme determining the range of investigations of neighbouring elements of metric during the percolation process in Lagrangian space. We consider a pixel ${\boldsymbol P}$ in Eulerian space, that we know to be already close to an element of metric $(i_{\rm m},j_{\rm m})$ because it is a direct neighbour of a pixel known to be in the region of influence of $(i_{\rm m},j_{\rm m})$. We remap it to its Lagrangian counterpart ${\boldsymbol Q}$ using the element of metric $(i_{\rm m},j_{\rm m})$. Then we aim to find the actually closest element of metric to it. To do so and speed up the process, four cases are considered, to take into account the small errors on ${\boldsymbol Q}$ (in particular the fact that the reconstructed ${\boldsymbol Q}$ depends on the element of metric under consideration, see Fig.~\ref{fig:recons}), that we write ${\boldsymbol E}_{\rm map}=(E_x,E_v)$. In case 4, ${\boldsymbol Q}$ is close enough to the element of metric: we do not need to investigate any neighbour and decide it to belong to the region of influence of $(i_{\rm m},j_{\rm m})$. In cases 2 and 3, we need only to investigate another element of metric, e.g. $(i_{\rm m},j_{\rm m}+1)$ in the upper part of the figure to decide if it closer to its own estimate of ${\boldsymbol Q}$ using the distance defined in equation (\ref{eq:distdef}). Finally case 1 is the most costly since we have to investigate three neighbouring element of metric, but it is seldom considered.}
\label{fig:ranges_nearest}
\end{figure}

To speed up the process, which is not really relevant here in 2D phase-space but crucial in higher number of dimensions, we reduce the range of investigations of values of $i_{\rm m}'$ to $[i_{\rm inf},i_{\rm sup}]$ where 
\begin{eqnarray}
\delta q_x < -\alpha \Delta x_{\rm m} &:& i_{\rm inf}=-1 , i_{\rm sup}=0, \label{eq:r1} \\
-\alpha \Delta x_{\rm m} \leq \delta q_x<  \alpha \Delta x_{\rm m}  &:& i_{\rm inf}=0 , i_{\rm sup}=0, \label{eq:r2} \\
\alpha \Delta x_{\rm m} \leq \delta q_x &:& i_{\rm inf}=0, i_{\rm sup}=1, \label{eq:r3}
\end{eqnarray}
with the dependence on $(i,j,i_{\rm m},j_{\rm m})$ of $\delta q_x$ still implicit and $\alpha$ is a parameter close to $1/2$. Similar conditions are set to define an interval $[j_{\rm inf},j_{\rm sup}]$. The parameter $\alpha$ is introduced to take into account the uncertainty ${\boldsymbol E}_{\rm map}=(E_x,E_v)$ on the remapping (\ref{eq:QofP}). This is illustrated by Fig.~\ref{fig:ranges_nearest}.  
However, even though we can actually estimate ${\boldsymbol E}_{\rm map}$ as described below in \S~\ref{sec:estierr}, we {\em do not use it} to calculate $\alpha$ which is fixed from the beginning. Indeed, our implementation favours performance over perfect accuracy, to make a future extension to 6-dimensional phase-space possible: while it is better to find in Lagrangian space the actual nearest element of metric to a test particle, this is not a must and the algorithm can still work without it, at the cost of some loss of accuracy on the long run. 

\begin{figure}
\includegraphics[width=\textwidth]{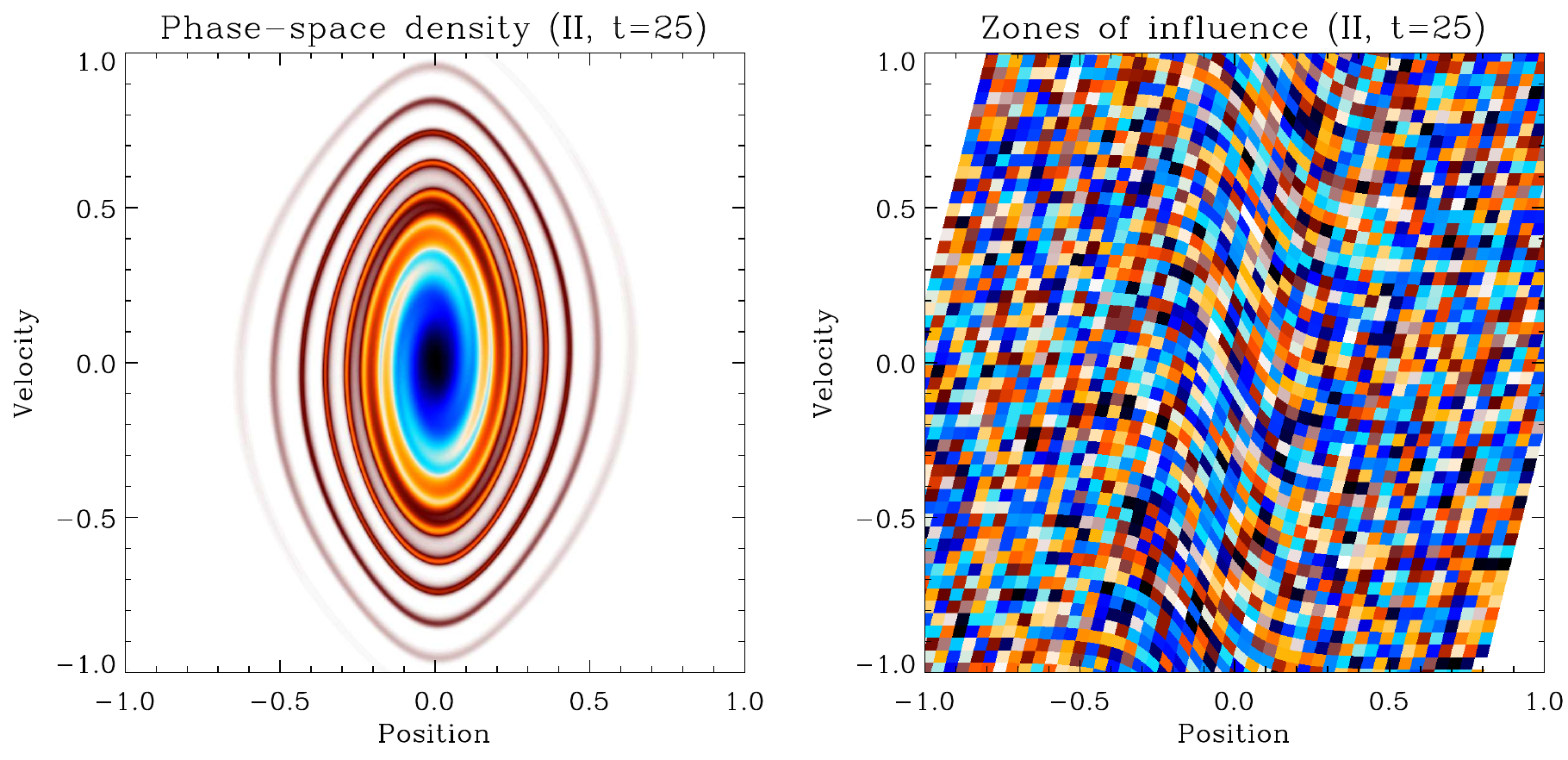}
\caption[]{Gaussian run and percolation. {\em Left panel:} the phase-space distribution function is shown at $t=25$ for a simulation with Gaussian initial conditions and parameters corresponding to item II of Table~\ref{tab:tab} in \S~\ref{sec:numexp}. {\em Right panel:} the corresponding Lagrangian zone of influence of each element of metric obtained from the percolation process is drawn at the end of a cycle (just before full resampling with new metric elements). To visualize the results more clearly, a random color has been associated to each element of metric.}
\label{fig:percoclo}
\end{figure}

To illustrate the result of the percolation process, Figure~\ref{fig:percoclo} displays, at $t=25$, the region of influence of each element of metric at the end of a cycle (i.e. just before resampling with new elements of metric) for the simulation II with Gaussian initial conditions described in \S~\ref{sec:numexp}. 

\subsubsection{Interpolation of the Lagrangian position of a test particle}
We now explain how the Lagrangian position of a test particle can be calculated in a smooth way with proper interpolation in Lagrangian space between predicates given by each of neighbouring element of metric. We define the following mapping function ${\boldsymbol Q}_{\rm r}({\boldsymbol P})$ from Eulerian to Lagrangian space
\begin{eqnarray}
{\boldsymbol Q}_{\rm r}({\boldsymbol P}) \equiv \frac{\sum_{i_{\rm m}',j_{\rm m}'} W\{2\,d[\delta{\boldsymbol Q}({\boldsymbol P},i_{\rm m}',j_{\rm m}')]\}\,{\boldsymbol Q}({\boldsymbol P},i_{\rm m}',j_{\rm m}')}{\sum_{i_{\rm m}',j_{\rm m}'} W\{2\,d(\delta{\boldsymbol Q}({\boldsymbol P},i_{\rm m}',j_{\rm m}')]\}},
\label{eq:interpol}
\end{eqnarray}
where function ${\boldsymbol Q}({\boldsymbol P},i_{\rm m},j_{\rm m})$ is the obvious extension to the continuum of its discrete counterpart ${\boldsymbol Q}(i,j,i_{\rm m},j_{\rm m})$ defined previously,  likewise for $\delta{\boldsymbol Q}({\boldsymbol P},i_{\rm m},j_{\rm m})$, and $W(x)$ is a smooth isotropic kernel function compactly supported. Clearly, function ${\boldsymbol Q}_{\rm r}({\boldsymbol P}) $ as defined above is at least continuous. Our choice for $W(x)$ is the spline of \citet[][]{Monaghan1992}:
\begin{equation}
W(x) \equiv \left\{ \begin{array}{ll} 
\displaystyle 1-\frac{3}{2} x^2 + \frac{3}{4} x^3, & x \leq 1, \\
\displaystyle \frac{1}{4} (2-x)^3, & 1 < x \leq 2,\\
\displaystyle 0, &  x > 2, \end{array}\right.
\end{equation}
which preserves the second order nature of the Taylor expansion around each metric element position. Our choice of $W(x)$ allows us to restrict the range of investigation of values of $i_{\rm m}'$ and $j_{\rm m}'$ to the respective intervals $[i_{\rm m}-1,i_{\rm m}+1]$ and $[j_{\rm m}-1,j_{\rm m}+1]$ where $(i_{\rm m},j_{\rm m})$ corresponds to the closest element of metric to point ${\boldsymbol P}$ as computed in previous paragraph, \S~\ref{sec:perco}, for sites $(i,j)$ of a grid.\footnote{Obviously, we assume here that the errors in the remapping (\ref{eq:QofP}) are smaller than the inter-element of metric separation.} Here, we indeed only need to consider points ${\boldsymbol P}=(\Delta x\, i + x_{\rm mid}, \Delta v\, j + v_{\rm mid})$ associated to such an Eulerian grid. To reduce furthermore the number of elements of metric investigated to perform the interpolation (\ref{eq:interpol})  while still taking into account small variations related to the previously discussed uncertainty ${\boldsymbol E}_{\rm map}$ on the calculation of ${\boldsymbol Q}$, we restrict the search to the intervals $[i_{\rm min},i_{\rm max}]$ and $[j_{\rm min},j_{\rm max}]$ defined as follow, with the same notations as in equations (\ref{eq:r1}), (\ref{eq:r2}) and (\ref{eq:r3}):
\begin{eqnarray}
\delta q_x < -\beta \Delta x_{\rm m} &:& i_{\rm min}=-1 , i_{\rm max}=0, \\
-\beta \Delta x_{\rm m} \leq \delta q_x<  \beta \Delta x_{\rm m}  &:& i_{\rm min}=-1 , i_{\rm max}=1, \\
\beta \Delta x_{\rm m}\leq \delta q_x &:& i_{\rm min}=0, i_{\rm max}=1,
\end{eqnarray}
where $\beta$ is a small parameter accounting for the error ${\boldsymbol E}_{\rm map}$,  similarly for $j_{\rm min}$ and $j_{\rm max}$,  with the additional constraints $i'_{\rm m} j'_{\rm m}=0$ if $i_{\rm min} i_{\rm max}=-1$ or $j_{\rm min} j_{\rm max}=-1$ to gain even more time. 

Again, similarly as stated at the end of \S~\ref{sec:perco}, the above algorithm is not guaranteed to provide a smooth interpolation if the error on the remapping associated to the metric elements becomes larger than some threshold. However, this should have little consequence on the algorithm accuracy and the gain of speed in high dimensional phase-space is considerable with such a procedure. Indeed, if ${\boldsymbol E}_{\rm map}$ becomes too large, the loss of accuracy on the characteristics themselves is much more dramatic than small discontinuities introduced on their interpolation.

\subsubsection{Estimates of errors on the Lagrangian position and choice of the margin parameters $\alpha$ and $\beta$}
\label{sec:estierr}
\begin{figure}
\begin{center}
\includegraphics[width=0.6\linewidth]{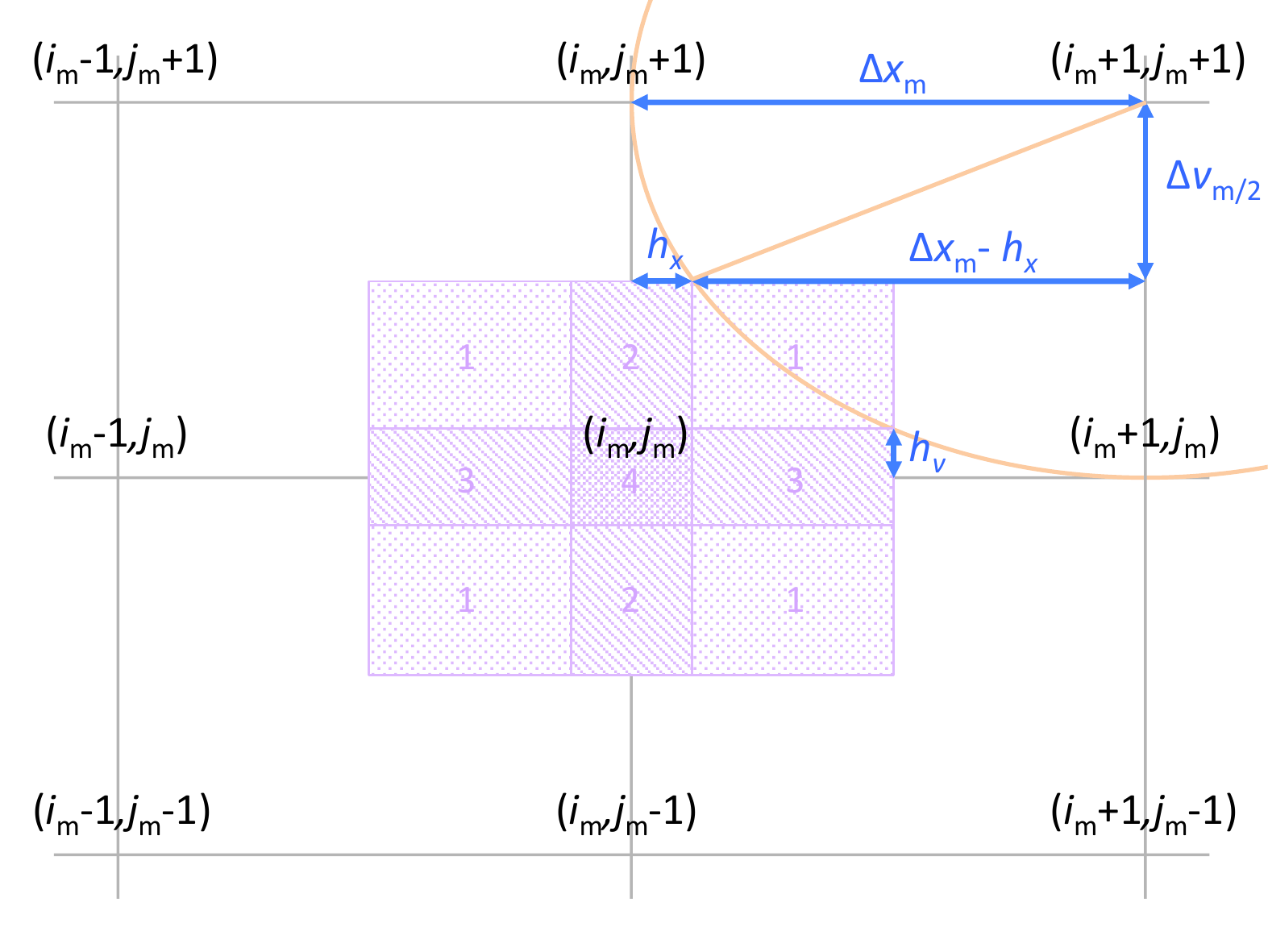}
\end{center}
\caption[]{Scheme determining the number of neighbouring elements of metric to consider for the interpolation (\ref{eq:interpol}) used to reconstruct the Lagrangian coordinate ${\boldsymbol Q}$ of a test particle ${\boldsymbol P}$ in the region of influence of metric element $(i_{\rm m},j_{\rm m})$. This figure also justifies equation (\ref{eq:beta}). Each element of metric has an ellipsoid region of influence (ocher ellipse) for computing the weights in equation (\ref{eq:interpol}). Outside this area, the element of metric does not contribute. Therefore, in the absence of uncertainty on the reconstruction of the Lagrangian coordinate, one can safely define 4 kinds of regions, as shown on the figure, with $h_x=(1-\sqrt{3}/2)\Delta x_{\rm m}$ and $h_v=(1-\sqrt{3}/2) \Delta v_{\rm m}$, steaming from the ellipse equations $(1-h_x/\Delta x_{\rm m})^2+1/4=1$ and $1/4+(1-h_v/\Delta v_{\rm m})^2=1$. In region 4, only the cross composed of $(i_{\rm m}-1,j_{\rm m})$, $(i_{\rm m}+1,j_{\rm m})$, $(i_{\rm m},j_{\rm m}-1)$, $(i_{\rm m},j_{\rm m}+1)$ and $(i_{\rm m},j_{\rm m})$ contributes to the interpolation (\ref{eq:interpol}). In regions 2 and 3, only 4 elements of metric contribute, namely $(i_{\rm m}-1,j_{\rm m})$, $(i_{\rm m}+1,j_{\rm m})$, $(i_{\rm m},j_{\rm m}+1)$ and $(i_{\rm m},j_{\rm m})$ for the upper area 2. In regions labelled by 1, four elements of metric contribute, e.g. $(i_{\rm m},j_{\rm m})$, $(i_{\rm m}+1,j_{\rm m})$, $(i_{\rm m},j_{\rm m}+1)$, $(i_{\rm m}+1,j_{\rm m}+1)$ in the upper right quadrant. In the presence of uncertainty,  we need to change the extensions of these regions to take it into account, as discussed further in the main text.}
\label{fig:myintfig}
\end{figure}

In the two subsections above, we defined two parameters $\alpha$ and $\beta$ but did not provide any numerical value. In principle these parameters could be user-defined, but we chose to fix them:
\begin{eqnarray}
\alpha&=&\frac{\sqrt{3}}{4}  \simeq 0.43,  \label{eq:monalpha}\\
\beta&=&\frac{1}{2} \left( 1-\frac{\sqrt{3}}{2} \right)  \simeq 0.067. 
\end{eqnarray}
The idea behind this choice is the following. Firstly, according to Figure~\ref{fig:myintfig}, since we do not want to consider any diagonal element of metric if $i_{\rm min} i_{\rm max}=-1$ or $j_{\rm min} j_{\rm max}=1$, we must have for full consistency,
\begin{equation}
\beta \leq 1-\frac{\sqrt{3}}{2}  \simeq 0.13. 
\label{eq:beta}
\end{equation}
Because there is some error on the reconstructed value of ${\boldsymbol Q}$, we take half this value, assuming that 
\begin{equation}
E_x/\Delta x_{\rm m}, E_y/\Delta v_{\rm m} \leq E_{\rm max} \equiv\frac{1}{2} \left( 1-\frac{\sqrt{3}}{2} \right)  \simeq 0.067. 
\end{equation}
Hence $\alpha=1/2-E_{\rm max}$ is given by equation (\ref{eq:monalpha}) to take into account this upper bound for the errors. With this setting, the algorithm is fully consistent only if the uncertainty on the remapped trajectories remains small enough, but even if $E_x/\Delta x_{\rm m}$ or $E_v/\Delta v_{\rm m}$ are larger than $E_{\rm max}$, it still works. Simply function ${\boldsymbol Q}_{\rm r}({\boldsymbol P})$ is not guaranteed to be completely smooth anymore. 

Note that in our algorithm, a global error $E_{\rm map}=\sqrt{(E_x/\Delta x_{\rm m})^2 +(E_v/\Delta v_{\rm m})^2}$ is estimated during the percolation process by comparing the value of ${\boldsymbol Q}$ proposed by a neighbouring element of metric to the value of ${\boldsymbol Q}$ proposed by the element of metric under consideration using the Lagrangian distance $d$ defined in equation (\ref{eq:distdef}). Such an error can in principle be used to force global remapping if $E_{\rm map}$ exceeds some threshold, for instance $E_{\rm max}$ to preserve full smoothness of function ${\boldsymbol Q}_{\rm r}(\boldsymbol{P})$. While this is an option in our code, we do not use it for the tests performed in \S~\ref{sec:numexp}. Indeed, we found difficult to find some reliable criterion on $E_{\rm map}$ and preferred to keep the number $n_{\rm s}$ of time steps between full resampling fixed and see how the results depend on the choice of $n_{\rm s}$, being aware of the fact that this is potentially suboptimal. Note also that $E_x$ and $E_v$ could be estimated as the residuals from the third order description of the local metric elements, but this is very costly in highly dimensional phase-space. 
\subsection{Equations of motion}
\label{sec:moti}
In $2D$-dimensional phase-space, the equations characterising the evolution of an element of metric with time are given by the following expressions with some obvious new notations:
\begin{eqnarray}
\frac{\partial {\boldsymbol x}}{\partial t}&=&{\boldsymbol v},\\
\frac{\partial \boldsymbol v}{\partial t}&=&{\boldsymbol a},\\
\frac{\partial T_{i,j}}{\partial t}&=& T_{i+D,j}, \quad i \leq D,\\
\frac{\partial T_{i+D,j}}{\partial t}&=& \sum_{r=1}^D \frac{\partial a_{i}}{\partial x_r} T_{r,j},\\
\frac{\partial H_{i,j,k}}{\partial t}&=&H_{i+D,j,k}, \quad i \leq D, \\
\frac{\partial H_{i+D,j,k}}{\partial t}&=&\sum_{r=1}^D \frac{\partial a_{i}}{\partial x_r} H_{r,j,k} +\sum_{r,m=1}^D \frac{\partial^2 a_{i}}{\partial x_r\,\partial x_m} T_{r,j} T_{m,k},
\end{eqnarray}
where ${\boldsymbol a}$ is the acceleration calculated by solving Poisson equation. Note thus that we need to have access at all times to the gradient and the Hessian of the gravitational force. In these equations we have dropped the ``m'' subscript for simplicity. 

The actual time step implementation uses a splitting algorithm with a drift phase during half a time step where only spatial positions are modified, followed with a kick phase where only velocities are modified during a full time step using the acceleration computed after the drift, followed again by a drift during half a time step, where spatial positions are modified using the new velocities. 

The first drift step or predictor is thus written
\begin{eqnarray}
  {\boldsymbol x}(t+\Delta t/2) &=& {\boldsymbol x}(t)+\frac{1}{2} {\boldsymbol v}(t) \Delta t,\\
  T_{i,j}(t+\Delta t/2) &=&T_{i,j}(t)+\frac{1}{2} T_{i+D,j}(t) \Delta t,\quad i \leq D,\\
  H_{i,j,k}(t+\Delta t/2)&=&H_{i,j,k}(t)+\frac{1}{2} H_{i+D,j,k} \Delta t, \quad i \leq D,\nonumber \\
&&
\end{eqnarray}
after which Poisson equation is solved to compute the acceleration ${\boldsymbol a}(t+\Delta t/2)$, its gradient and its Hessian. To compute the acceleration, we first need to estimate the phase-space distribution function $f(\boldsymbol x,\boldsymbol v,t+\Delta t/2)$ on the Eulerian grid. To map $f$ back to initial conditions (or previous stage after full remapping), we use the elements of metric as transformed by the predictor step. 

The kick phase reads
\begin{eqnarray}
 {\boldsymbol v}(t+\Delta t) &=& {\boldsymbol v}(t)+{\boldsymbol a}(t+\Delta t/2) \Delta t, \\
  T_{i+D,j}(t+\Delta t) &=&T_{i,j}(t)+\sum_{r=1}^D \frac{\partial a_{i}}{\partial x_r} T_{r,j}(t+\Delta t/2) \,\Delta t, \nonumber\\
& & \\
H_{i+D,j,k}(t+\Delta t)&=&H_{i+D,j,k}(t) +\sum_{r=1}^D \frac{\partial a_{i}}{\partial x_r} H_{r,j,k}(t+\Delta t/2)\, \Delta t\nonumber \\
 && +\sum_{r,m=1}^D \frac{\partial^2 a_{i}}{\partial x_r\,\partial x_m}\, T_{r,j}(t+\Delta t/2) \,T_{m,k}(t+\Delta t/2)\, \Delta t.
\end{eqnarray}

Finally the second drift phase, or corrector, is expressed as follows:
\begin{eqnarray}
  {\boldsymbol x}(t+\Delta t) &=& {\boldsymbol x}(t+\Delta t/2)+\frac{1}{2} {\boldsymbol v}(t+\Delta t) \Delta t,\\
  T_{i,j}(t+\Delta t) &=&T_{i,j}(t+\Delta t/2)+\frac{1}{2} T_{i+D,j}(t+\Delta t) \Delta t, \quad i \leq D,\\
  H_{i,j,k}(t+\Delta t)&=&H_{i,j,k}(t+\Delta t/2)+\frac{1}{2} H_{i+D,j,k}(t+\Delta t)\Delta t, \quad i \leq D.
\end{eqnarray}

With our predictor-corrector scheme, it is possible to adopt a slowly variable time step.  In this case, we use the following dynamical constraint \citep[see, e.g.,][]{Alard2005,Colombi2014},
\begin{equation}
\Delta t \leq \frac{C_{\rm dyn}}{\sqrt{\rho_{\rm max}}}, \quad C_{\rm dyn} \ll 1,
\end{equation}
where $\rho_{\rm max}={\rm max}_x \rho(x)$ is the maximum value of the projected density, $\rho(x) \equiv \int f(x,v) {\rm d}v$. 
Note that for the tests performed in \S~\ref{sec:numexp}, we used a constant time step, but this should not have critical effects on the numerical results.

\subsection{Calculation of the force and its derivatives}
\label{sec:forc}
In practice, the force is calculated on a grid and its gradient and Hessian estimated by simple finite difference methods. 

In our 2D phase-space implementation, the acceleration is calculated at discrete points $i$ with coordinate $x_i=x_{\rm mid}+\Delta x\,i$ corresponding to the projection of the Eulerian phase-space grid. To estimate it, we first calculate the projected mass $M(i)$ inside each pixel $i$ by summing up the phase-space density calculated on each point $(i,j)$ of the Eulerian phase-space grid. In the present work we consider isolated systems in an empty space, which means that the acceleration at a given point $x$ is simply proportional to the total mass at the right of point $x$ minus the total mass at the left of $x$. This implies, within some trivial factor depending on code units
\begin{equation}
a(i)=M_{\rm total}-M_{\rm left}(i)-M_{\rm left}(i-1),
\label{eq:acccomp}
\end{equation}
with
\begin{equation}
M_{\rm left}(i)=\sum_{m \leq i} M(m),
\end{equation}
and where $M_{\rm total}$ is the total mass of the system. Note thus that the mass in pixel $i$ does not contribute to $a(i)$ as it is supposed to be distributed symmetrically inside it.\footnote{{Note that a potentially better procedure could consist in using a staggered mesh for computing the acceleration, in which the nodes would correspond to boundaries of each pixel of the grid. }} In larger number of dimensions, Poisson equation is more complex to solve but this can be done on a fixed grid resolution with FFT methods, whether the system is spatially periodic or isolated as supposed here. 

Our estimate of the first and second derivatives of the acceleration at grid points is written
\begin{eqnarray}
\frac{\partial a}{\partial x_i} &=&\frac{a(i+1)-a(i-1)}{\Delta x},\\
\frac{\partial^2 a}{\partial x_i^2}&=&\frac{a(i+1)+a(i-1)-2 a(i)}{(\Delta x)^2},
\end{eqnarray}
 which is equivalent to fit the acceleration with a second order polynomial on the three successive points corresponding to $i-1$, $i$ and $i+1$. 

To compute the force at an arbitrary point of the computing domain, we perform dual TSC \citep[triangular shaped cloud, see e.g.,][]{Hockney1988} interpolation from the grid, i.e., 
\begin{eqnarray}
a(x)&=&\frac{1}{2} \left( \frac{1}{2} - w \right)^2 a(i-1) +\left( \frac{3}{4} - w^2\right) a(i)+\frac{1}{2} \left( \frac{1}{2}+w \right)^2 a(i+1)\\
i&=&\lfloor (x-x_{\rm mid})/\Delta x \rceil\\
w&=& (x-x_{\rm mid})/\Delta x-i
\end{eqnarray}
and likewise for its gradient and its second derivative. Note that we also tested linear interpolation, which leads in practice to nearly the same results as TSC for all the numerical tests we discuss in next section.

Clearly, the way we estimate the force and its successive derivatives is probably too naive but actually works pretty well for all the experiments we did so we did not bother to try to improve this part of our algorithm. However, this should be investigated in a future work. 

\section{Numerical experiments}
\label{sec:numexp}

\begin{table}

\begin{center}
\begin{tabular}{c|l|l|c|c}
\hline
Designation & $\Delta$ & $\Delta_{\rm m}$ & $n_{\rm s}$ & $t_{\rm CPU}$ \\
\hline
I    & 0.002 & 0.02 & 25 & 1.0\\
II   & 0.002 & 0.04 & 25 &  1.0 \\
III  & 0.002 & 0.01 & 25 & 1.2 \\
IV  & 0.002 & 0.005 & 25 & 1.4 \\
V   & 0.002 & 0.02 & 100 & 1.1 \\
VI  & 0.002 & 0.02 & 50 & 1.1 \\
VII  & 0.002 & 0.02 & 10 & 1.2 \\
VIII & 0.002 & 0.02 & 5 & 1.5 \\
IX  & 0.002 & 0.02 & 1  & 3.2\\
X  & 0.001 & 0.01 & 25 & 4.0\\
XI & 0.004 & 0.04 & 25 & 0.4 \\
\hline
XII & 0.0005 & \ \ --   & -- & 8.8\\
XIII & 0.00707 & \ \ -- & -- & 4.5\\
XIV & 0.001 & \ \ -- & --  & 2.3 \\
XV & 0.00141 &\ \ -- & -- & 1.2 \\
XVI & 0.002 & \ \ -- & -- & 0.7
\end{tabular}
\end{center}

\caption[]{Parameters of the simulations performed in this work, namely the grid cell size, $\Delta=\Delta x=\Delta y$, the initial separation between each metric element, $\Delta_{\rm m}=\Delta x_{\rm m}=\Delta y_{\rm m}$, the number of sub-cycles, $n_{\rm s}$, between resamplings { and the approximate total CPU time expressed in units of the time spent for simulation I.  The calculation of $t_{\rm CPU}$ was performed for the simulations with initial conditions corresponding to the random set of stationary clouds as described in \S~\ref{sec:comp}, but the results do not change drastically for other initial conditions. A horizontal line separates the runs performed with {\tt Vlamet} (I--XI) from those performed with the standard splitting algorithm (XII--XVI). } Additionally, the time step $\Delta t$ was chosen constant, given by $\Delta t=0.01$, $0.01$ and $0.0025$ for the stationary, Gaussian and random set of halos, respectively. The two last choices come from the constraints on the time step obtained in the waterbag runs by \citet[][]{Colombi2014}. }
\label{tab:tab}
\end{table}
 To test the method, we performed an ensemble of simulations of which the parameters are listed in Table~\ref{tab:tab}. In \S~\ref{sec:comp}, we compare the results obtained with {\tt Vlamet} to the standard splitting algorithm of \citet{Cheng1976} with cubic B-spline interpolation as well as the extremely accurate entropy conserving waterbag simulations of \citet[][hereafter CT14]{Colombi2014}. Before going into the details of these comparisons, we first provide in \S~\ref{sec:perf} additional technical information on the performances of {\tt Vlamet} and of the splitting algorithm.\footnote{{Comparisons to the waterbag code in terms of performances are irrelevant here, since the future goal is to generalize {\tt Vlamet} to higher number of dimensions in phase-space. The waterbag method, extremely accurate but already costly in 2-dimensional phase-space, becomes prohibitive in higher number of dimensions. For instance, in 4-dimensional phase-space, it consists in following, in an adaptive fashion, ensembles of hypersurfaces sampled with tetrahedra \citep[see, e.g.][]{Sousbie2016}. }} In particular, we discuss parallelisation issues as well as extension to higher number of dimensions.

\subsection{Performances: parallel scaling and computing cost}
\label{sec:perf}

\begin{figure}
\begin{center}
\includegraphics[width=0.5 \textwidth]{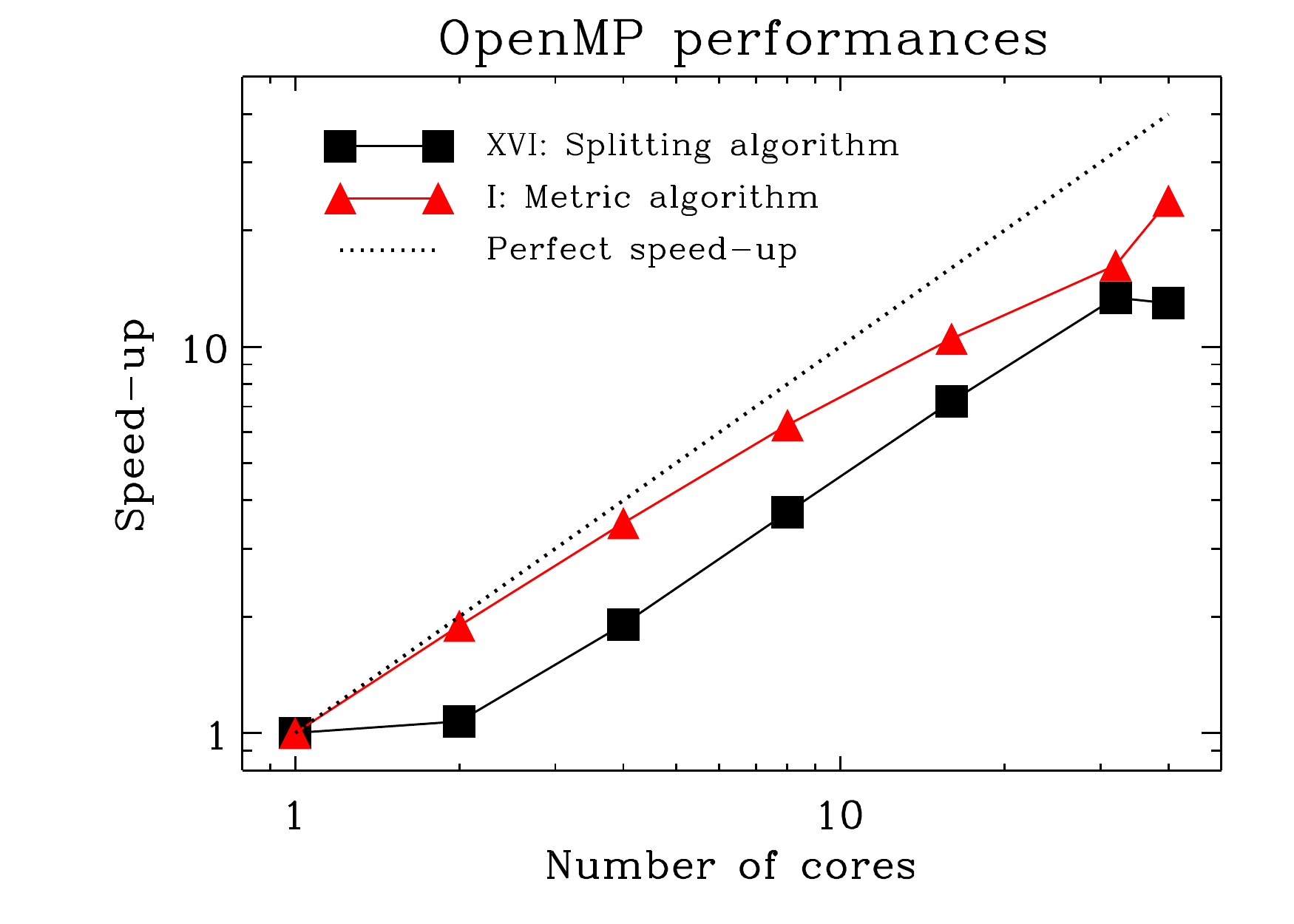}
\end{center}
\caption[]{{Accelerations of our implementations of {\tt Vlamet} (in red) and of the splitting algorithm (in black) as functions of the number of cores on a shared memory architecture with OpenMP library. The plot is performed for the parameters corresponding to runs I and XVI in Table~\ref{tab:tab}, starting from an ensemble of stationary clouds. Perfect speedup is represented as a dotted line. }}
\label{fig:openMP}
\end{figure}

Both {\tt Vlamet} and the splitting algorithm are parallelised on shared memory architecture using OpenMP library, but no attempt has been made to optimize cash memory, so there is still room for improvement. Parallel scaling is displayed on Fig.~\ref{fig:openMP}, which shows that the two codes present a reasonable speedup, greater than 10 on 32 cores. The scaling is slightly better for the metric code than for the splitting algorithm but this comparison is irrelevant without further optimisation. The main point here is to show that OpenMP parallelisation is rather straightforward for {\tt Vlamet}, while it is already well known that both distributed and shared memory parallelisations are not a problem for the splitting algorithm \cite[see, e.g.,][]{Crouseilles2009}. 

As an additional note, but this is something we leave for future work, there is in principle no particular difficulty in parallelising {\tt Vlamet} on a distributed memory architecture with the MPI library. Except for solving Poisson equation and for cubic B-spline interpolation that we discuss just after, all the operations performed in the algorithm, including the percolation process, are of local nature, hence trivially compatible with domain decomposition. In 4 and 6-dimensional phase-space, introducing a k-d tree structure to focus on important regions of phase space is also quite standard in parallel programming. The solution to Poisson equation is well known in many parallel implementations. For instance there exist parallel Fast Fourier Transform libraries such as FFTW to compute the gravitational potential on a fixed resolution grid. As for spline interpolation, a domain decomposition implementation is possible using the method proposed by \citet{Crouseilles2009}. In this case, interpolation is localised to each domain by using Hermite boundary conditions between the domains with an ad hoc reconstruction of the derivatives. In this implementation, each domain would for instance correspond to a leaf of the k-d tree mentioned above. Of course, as long as we did not actually implement the algorithm on distributed memory architecture, this discussion stays at the speculative level. 

In terms of total computational time, interesting conclusions can be driven from examination of fifth column of Table~\ref{tab:tab}, which indicates, for each run, the approximate total CPU cost in units of simulation I. Neglecting the  Poisson equation resolution step, both {\tt Vlamet} and the splitting algorithm are expected in general to scale roughly like the number $n_{\rm cells} \propto \Delta^{-2D}$ of cells in the grid representing phase-space, where $D$ (here equal to unity) is the dimension of space. This is approximately the case in Table~\ref{tab:tab} if $\Delta$ is small enough. The cost also increases for {\tt Vlamet} when resamplings of the phase-space distribution function are more frequent (in practice, if $n_{\rm s} \la 10$ for the runs we did) or when the element of metric density increases (in practice, if $\Delta_{\rm m} \la 10 \Delta$ for the runs we did).

When it comes to really compare {\tt Vlamet} to the standard splitting algorithm, we focus on single thread runs, assuming that full parallelisation is equally optimal on both algorithms. We noticed that with the same spatial resolution $\Delta$, {\tt Vlamet} is about $\alpha_{\rm slower}=1.4-2$ times slower than the splitting algorithm for $n_{\rm s} \ga 25$ and $\Delta_{\rm m}/\Delta \ga 2.5$.  How these estimates generalize to 4 and 6 dimensions remains an open question, but we can try to infer some rough estimates. From now on we restrict to 6D phase-space. In this case, the critical step in {\tt Vlamet} is the cubic spline interpolation during the resampling phase, which is expected to be about 171 times more costly than a step of the standart splitting algorithm, as discussed in introduction of \S~\ref{sec:int}. The rest of the algorithm, on the other hand, should perform comparably well to the splitting scheme, i.e. should be $\alpha_{\rm loss}$ times slower with $\alpha_{\rm loss}$ of the order of unity. Hence we expect, in six dimensions, a factor $\alpha_{\rm slower}^{6D}\simeq (n_{\rm s} \times \alpha_{\rm loss} + 171)/n_{\rm s}$ between {\tt Vlamet} and the standard splitting algorithm, that is $\alpha_{\rm slower}^{6D}$ ranging from about 2 in the optimistic cases (e.g. $\alpha_{\rm loss} \simeq 1$ and $n_{\rm s}=200$) to about 20 in the pessimistic ones (e.g. $\alpha_{\rm loss}=10$ and $n_{\rm s}=15$). While a factor as large as 20 might seem potentially prohibitive, remind that our code is supposed to be less diffusive than the splitting algorithm, as we shall illustrate next, hence should require less spatial resolution. A simple factor 2 in resolution corresponding to a gain of about $2^6=64$ in speed and memory would make {\tt Vlamet} extremely competitive.

\subsection{Test runs: comparisons between codes}
\label{sec:comp}

Before going further, it is important here to specify the units in which we are solving Vlasov-Poisson equations. Our choice is the same as in CT14, i.e. we are studying the following system of equations:
\begin{eqnarray}
\frac{\partial f}{\partial t} + v \frac{\partial f}{\partial x} - \frac{\partial \phi}{\partial x} \frac{\partial f}{\partial v}=0\\
\frac{\partial^2 \phi}{\partial x^2} = 2 \rho=2 \int f(x,v,t) {\rm d} v.
\end{eqnarray}
To test {\tt Vlamet},  three kinds of initial conditions were considered, namely the thermal equilibrium solution, Gaussian initial conditions and a random set of stationary clouds, as detailed below. 
\begin{itemize}
\item {\em The thermal equilibrium solution} \citep{1942ApJ....95..329S,Camm1950,Rybicki1971} reads
 \begin{equation}
f_{\rm S}(x,v)=\frac{\rho_{\rm S}}{[{\rm ch}(\sqrt{\sqrt{2\pi} \rho_{\rm S}/\sigma_{\rm S}} x)]^2} \exp\left[ -\frac{1}{2} \left( \frac{v}{\sigma_{\rm S}}\right)^2\right].
\label{eq:apofs}
\end{equation}
In practice, we set
\begin{equation}
f_{\rm ini}(x,v)=f_{\rm S}(x,v) g_{\eta}[f_{\rm S}(x,v)],
\label{eq:apod}
\end{equation}
with the apodizing function $g_{\eta}$ given by
\begin{equation}
g_{\eta}(y)=\eta \max\left[ 1+2\ {\rm th}\left(\frac{y-\eta}{\eta}\right),0 \right]. 
\end{equation}
For the runs performed in this paper, we took $\rho_{\rm S}=4$, $\sigma_{\rm S}=0.2$ and $\eta=0.02$,  which means that the maximum projected density of the system is $\rho_{\rm max} \simeq 2$. The harmonic dynamical time corresponding to such a density is, in our units, 
\begin{equation}
T_{\rm dyn}=\frac{2\pi}{\sqrt{2\rho_{\rm max}}}.
\label{eq:dynT}
\end{equation}
The simulation was run up to $t=100$, which corresponds to $100/T_{\rm dyn} \simeq 32$ dynamical times.
Thermal equilibrium is stable and hence represents a crucial test for a code, which has to be able to maintain it accurately. Figure~\ref{fig:stationfxv} shows how {\tt Vlamet} performs for runs I, II, III and IV, when considering the difference $\delta f=f(x,v,t)-f_{\rm ini}(x,v)$ at $t=100$, i.e. checks the behaviour of the code when changing the distance $\Delta_{\rm m}$ between metric elements while keeping resolution fixed to $\Delta=0.002$. For comparison, a curve corresponding to the splitting algorithm run XVI is also shown at same resolution $\Delta$. As expected, {\tt Vlamet} performs very well, as long as $\Delta_{\rm m}/\Delta$ remains sufficiently small, $\Delta_{\rm m} /\Delta \la 10$, to avoid too large errors on the reconstruction of Lagrangian positions from the metric elements. We also checked that $|\delta f|$ decreases when augmenting the spatial resolution of the simulation. Note that the splitting algorithm seems to perform even better than {\tt Vlamet}, as suggested by the turquoise curve on right panel of Fig.~\ref{fig:stationfxv}.  A residual error remains in all cases, since there is a clear convergence to a non zero $\delta f$. From performing the a run similar to I but with a twice larger time step and another one with a truncature parameter $\eta$ twenty times smaller, we see that this small residual can be attributed to the finite value of the time step and especially the (rather strong) truncature of $f$ with apodizing function $g_{\eta}$, which implies that the initial conditions constructed this way do not correspond exactly to thermal equilibrium.  Although we do not have any rigorous mathematical argument to support the claim that follows, decreasing $\Delta t$ and the value of $\eta$ should make this residual tend to zero, both for {\tt Vlamet} and the splitting algorithm, provided that resolution $1/\Delta$ becomes also arbitrarily large and the ratio $\Delta_{\rm m}/\Delta$ is kept sufficiently small.s 
\begin{figure}
\includegraphics[width=\textwidth]{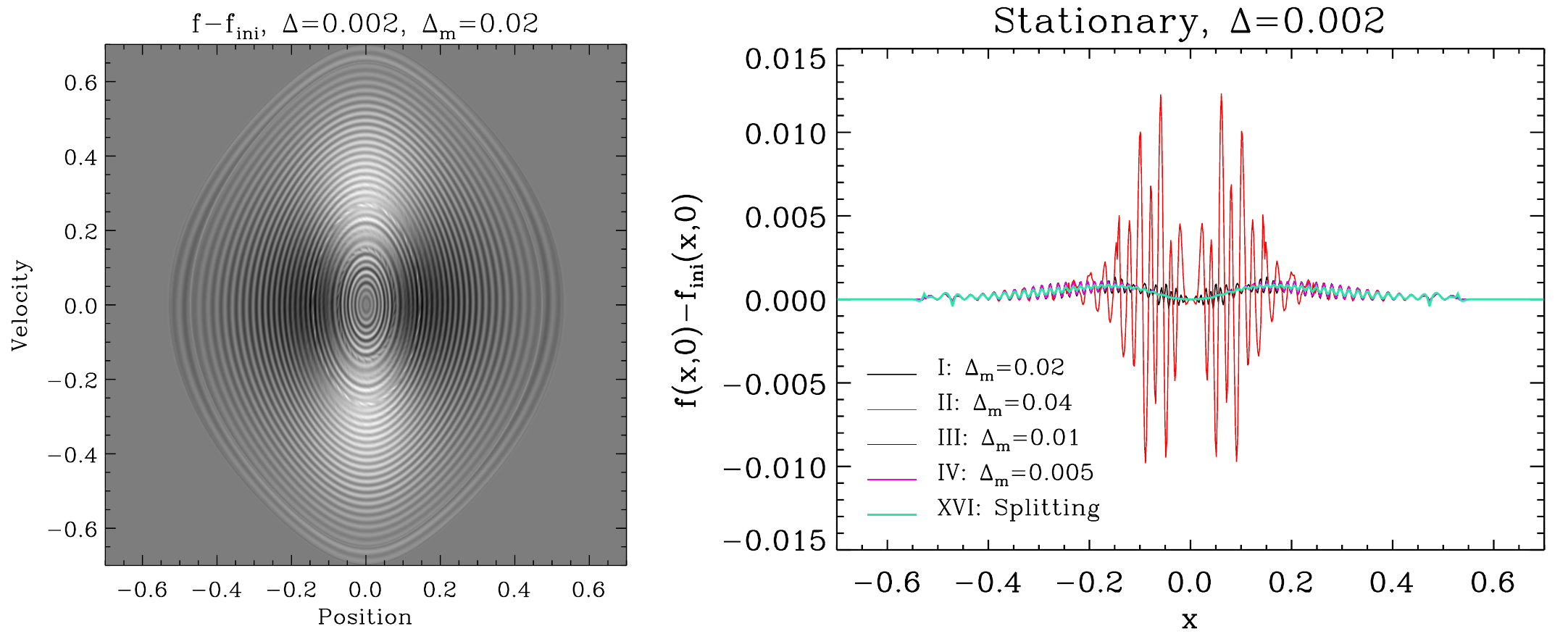}
\caption[]{{ Deviation from stationarity for simulations with $\Delta=0.002$. Left panel shows the difference $\delta f$ between the phase-space density measured at $t=100$ and the initial one for simulation I. Darker regions correspond to larger values of $\delta f$, which ranges in the interval $\delta f \in [-0.0014,0.0014]$. On right panel, the value of $\delta f(x,0)$ is plotted for simulations I, II, III, IV (i.e. varying inter-element of metric distance $\Delta_{\rm m}$ while keeping spatial resolution $\Delta=0.002$ fixed) and XVI. Note that the purple and blue curves nearly superpose on each other. }}
\label{fig:stationfxv}
\end{figure}
\item {\em Gaussian initial conditions} are given by
\begin{eqnarray}
f_{\rm ini}(x,v) & = & G(x,v), \quad x^2+v^2 \leq {\cal R}^2, \label{eq:monfg} \\
        & = &  G(x,v) \max\left[ 1+2\ {\rm th}\left( \frac{{\cal R}-\sqrt{x^2+v^2}}{\eta_{\rm G}}\right), 0 \right], \quad \ \ x^2+v^2 > {\cal R}^2, \label{eq:monfg2}
\end{eqnarray}
with
\begin{equation}
G(x,v)\equiv \rho_{\rm G} \exp\left( -\frac{1}{2} \frac{x^2+v^2}{\sigma_{\rm G}^2} \right). \label{eq:gauini}
\end{equation}
In practice, to compare the output of {\tt Vlamet} to the waterbag result of CT14, we take ${\cal R}=1$, $\rho_{\rm G}=4$, $\sigma_{\rm G}=0.2$ and $\eta_{\rm G}=0.02$, which makes the total mass of the system approximately equal to unity for a Gaussian truncated at 5 sigmas.  The measured maximum projected density of this system, is, after relaxation, $\rho_{\rm max}=2.95$, so simulating this system up to $t=100$ corresponds to following it during approximately 39 dynamical times according to equation (\ref{eq:dynT}).    Figure~\ref{fig:gaussianfxv} shows, at various times, the phase-space distribution function obtained from run I and the waterbag run {\tt Gaussian84} of  CT14. The agreement between both simulations seems visually close to perfect. A more careful examination of the details of the phase-space distribution function shows small deviations which can be reduced by augmenting the resolution of the {\tt Vlamet} run.  This is illustrated by Fig.~\ref{fig:gaussian_profile}, which displays on a small interval, the projected density for runs I, X, XI (different resolutions $\Delta$ but fixed $n_{\rm s}=25$ and fixed ratio $\Delta_{\rm m}/\Delta=10$) and {\tt Gaussian84}. What is particularly interesting in this figure are the turquoise curves corresponding to runs performed with the splitting algorithm, namely XII, XIV and XVI. We notice a very good mach between {\tt Vlamet} and the splitting algorithm but only if this latter is run at twice the resolution of {\tt Vlamet}!  This figure already demonstrates the significantly weaker level of diffusion of {\tt Vlamet} compared to the splitting algorithm. 
\begin{figure}
\includegraphics[width=\textwidth]{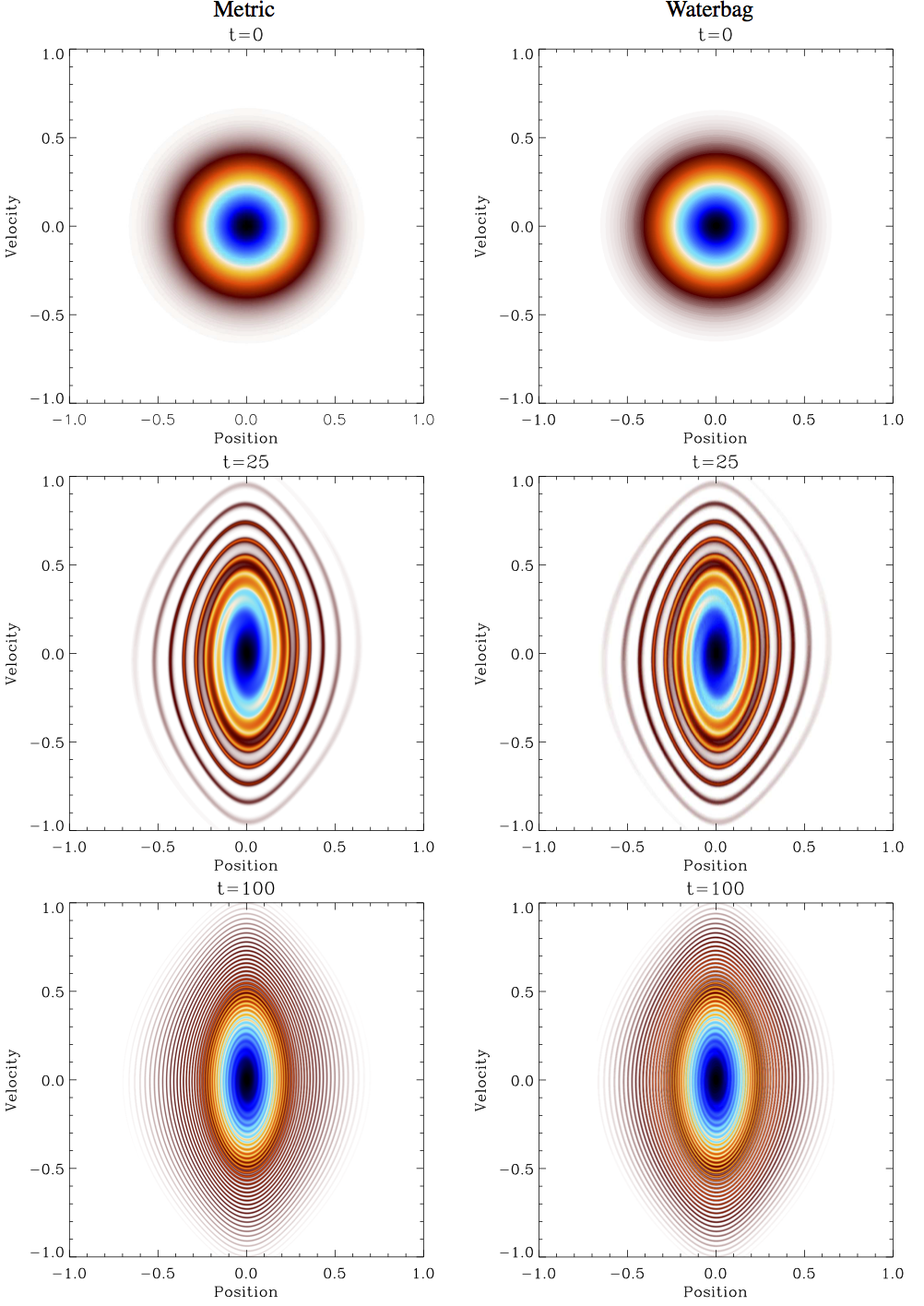}
\caption[]{Phase-space density in the simulations with Gaussian initial conditions. The simulation I with the metric method is compared to a waterbag run of \citet[][]{Colombi2014}. In the later case, the phase-space distribution function is represented with 84 waterbags.}
\label{fig:gaussianfxv}
\end{figure}
\begin{figure}
\includegraphics[width=\textwidth]{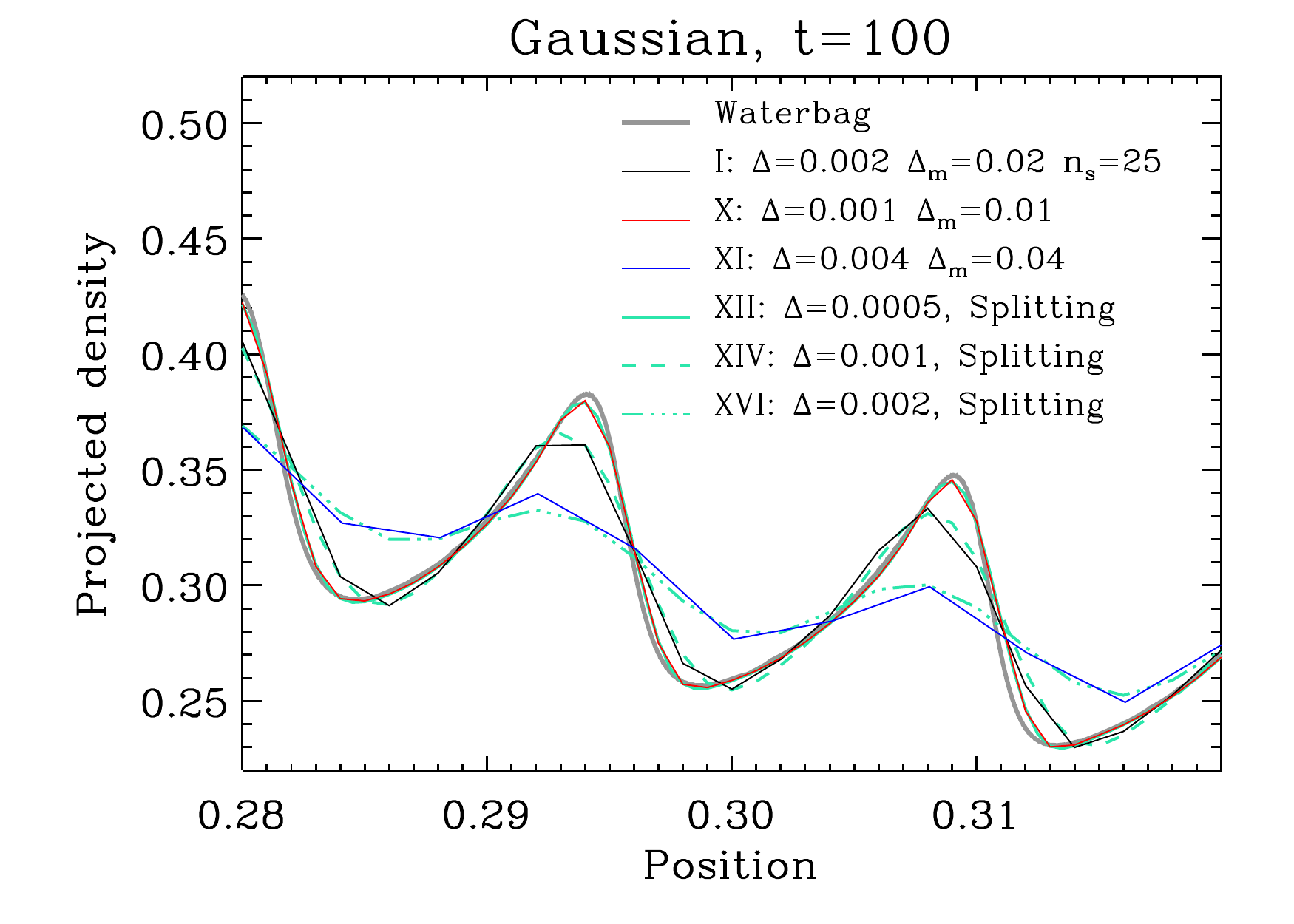}
\caption[]{{ Projected density in the simulations with Gaussian initial conditions: detail. The simulations performed with {\tt Vlamet}, I, X and XI are compared to the results obtained with the splitting algorithm, XII, XIV and XVI, as well as a waterbag run of \citet[][]{Colombi2014}, which is supposed to represent the ``exact'' solution. Note that while the waterbag measurements are displayed at the fine grained level, the results from the {\tt Vlamet} and the splitting algorithm runs are shown at the grid resolution, $\Delta$. }}
\label{fig:gaussian_profile}
\end{figure}
\item {\em The ensemble of stationary clouds}  corresponds to a set of halos following the stationary solution (\ref{eq:apofs}), created exactly as in CT14. In particular, their profile follows equation (\ref{eq:apofs}) with $\rho_{\rm S}=6$ and their velocity dispersion, $\sigma_{\rm S}$, ranges in the interval $[0.005,0.1]$. The components are added on the top of each other in phase-space, to obtain the desired initial distribution function $f_{\rm ini}(x,v)$, which is apodized as in equation (\ref{eq:apod}) with $\eta=0.05$. The interesting fact about this system is that it was found experimentally
by CT14 to be chaotic. In figure~\ref{fig:randomfxv},  the phase-space distribution function is shown at various times for simulation I and run {\tt Random} of CT14. At $t=25$,  the agreement between both simulations is impressive, despite the fact that the phase-space distribution function is sampled with only {\rm three} waterbags in the {\tt Random} run of CT14. At $t=80$, the system presents remarkable structures: some halos have been destroyed by tidal effects and a complex filamentary structure has grown. At this time, the {\tt Vlamet} and the waterbag runs diverge from each other, but remain still in agreement at the macroscopic level. The fact that the simulations diverge from each other is not surprising since the run performed by CT14 samples the phase-space distribution function with only 3 waterbags. The differences are also amplified by the intrinsic chaotic nature of the system. Note that the {\tt Vlamet} run is subject, at this point, to significant aliasing effects. For instance, the minimum value of the measured phase-space distribution function on lower-left panel of Fig.~\ref{fig:randomfxv} is equal to $f_{\rm min}=-0.38$, which is of rather large magnitude given the maximum theoretical value of $f$, $f_{\rm max}=6$. These aliasing effects are inherent to the spline interpolation procedure we are using. Importantly, aliasing effects do not affect the dynamical properties of the system: increasing the resolution or decreasing it (runs X and XI) does not change the results significantly, despite the chaotic nature of the system, except at very small scales close to spatial resolution, of course.  As illustrated below quantitatively, similarly as in the Gaussian case discussed above, simulations with the splitting algorithm provide comparable results to {\tt Vlamet} but only if performed at better resolution.
\begin{figure}
\includegraphics[width=\textwidth]{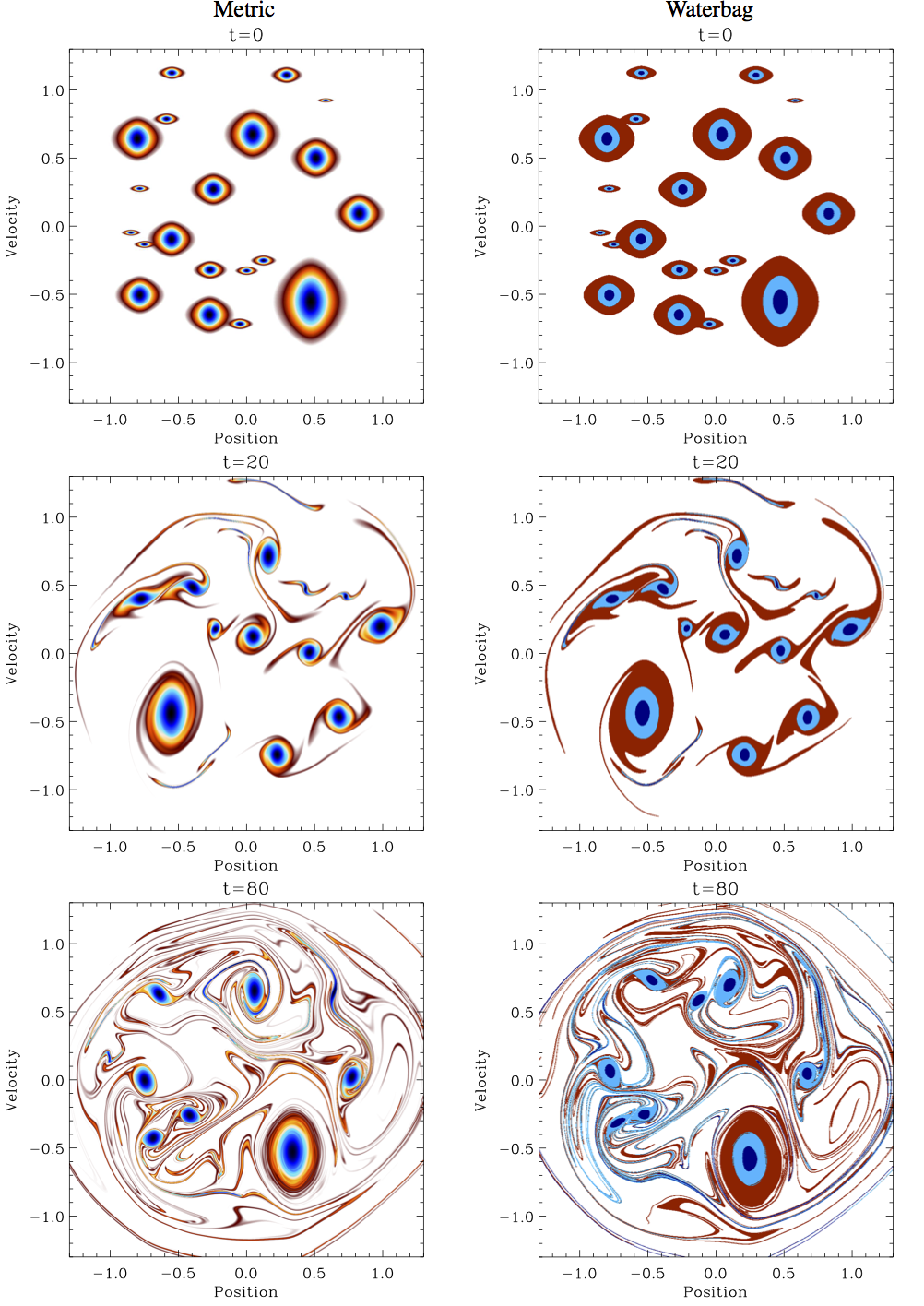}
\caption[]{Phase-space density in the simulations with random initial conditions. The simulation I with the metric method is compared to a waterbag run of \citet{Colombi2014}. In the later case, the phase-space distribution function is represented with only with three waterbag levels.}
\label{fig:randomfxv}
\end{figure}
\end{itemize}

 To perform more quantitative tests, we checked for conservation of total energy, which can be conveniently written
\begin{equation}
E=\frac{1}{2} \int v^2 f(x,v,t)\ {\rm d}x\ {\rm d}v-\frac{1}{4} \int \left[ a(x,t)^2 - M_{\rm total}^2 \right] {\rm d} x,
\end{equation}
where $a(x,t)$ is the gravitational force and $M_{\rm total}$ the total mass. These integral are computed directly as sums on the grid used to sample the phase-space distribution function and the force.  In addition, we check for conservation of the following Casimir
\begin{equation}
S=-\int f(x,v) \log|f(x,v)| {\rm d} x {\rm d} v,
\end{equation}
which is numerically estimated the same way as total energy. This quantity would reduce to Gibbs entropy if positivity of the phase-space distribution function was preserved. While we know it is not rigorously correct, we shall still designate $S$ by entropy. We also tested for total mass conservation, as well as $L^1$ and $L^2$ norms, but do not show the results here, to avoid superfluous content. Indeed, energy and entropy were found in practice to be sufficiently representative of the system when it comes to test the accuracy of  {\tt Vlamet}. 
 
\begin{figure}
\includegraphics[width=\textwidth]{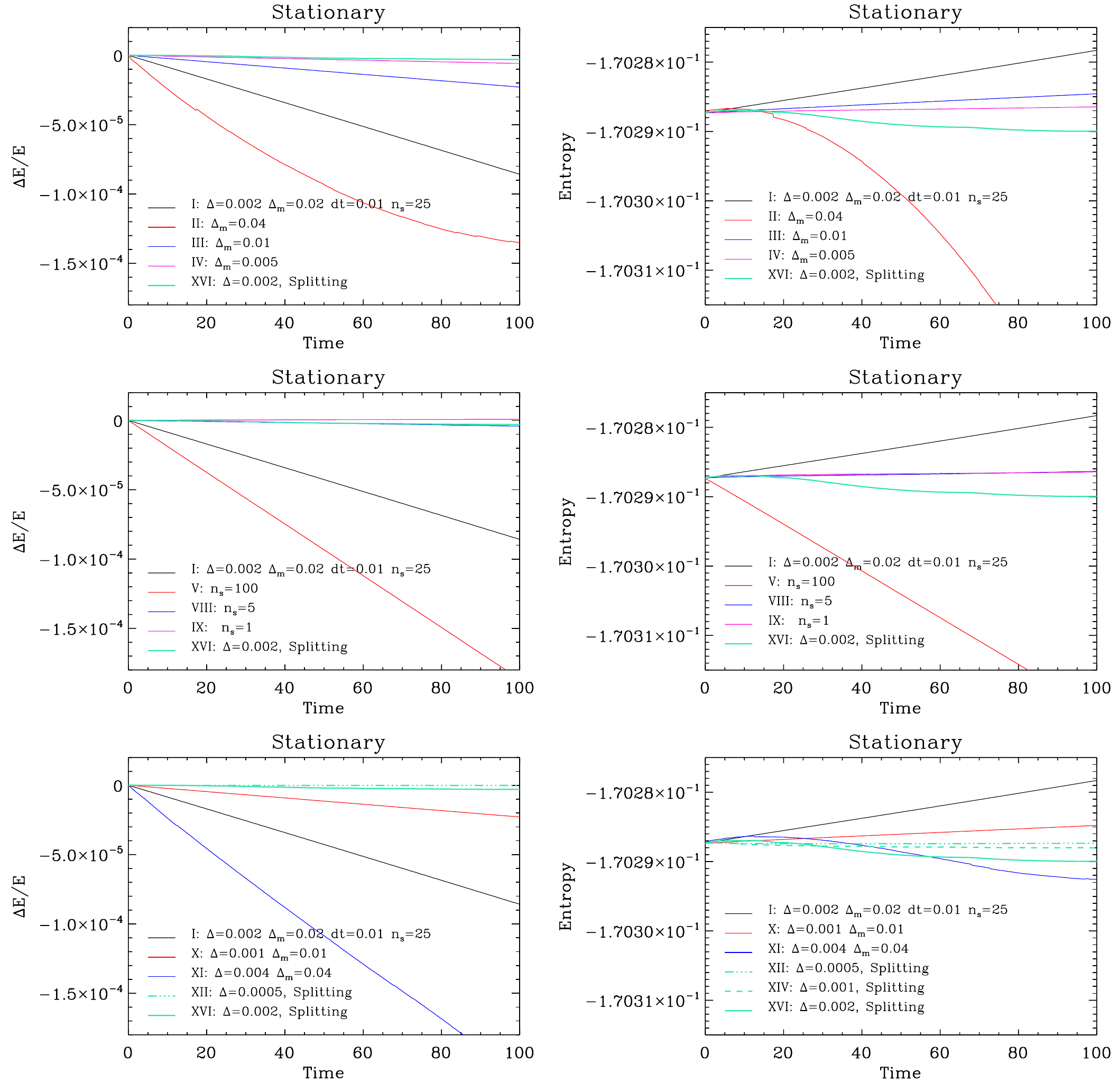}
\caption[]{Conservation of energy (left panels) and entropy (right panels) in the stationary runs. Each line of panels corresponds to varying one parameter of the simulations as listed in Table~\ref{tab:tab}. In the first line of panels, we vary the elements of metric sampling, $\Delta_{\rm m}=\Delta x_{\rm m}=\Delta v_{\rm m}$, while keeping other parameters, in particular spatial resolution $\Delta=\Delta x=\Delta v$, fixed. In the second line of panels, it is the number of time steps $n_{\rm s}$ between each resampling which changes. Finally, in the last line of panels, overall resolution is changed, i.e. both $\Delta$ and $\Delta_{\rm m}$ while keeping the ratio $\Delta_{\rm m}/\Delta$ fixed. { For comparison, results from the splitting algorithm are also displayed as turquoise curves as indicated on each panel. In the latter case, the only parameter of interest is spatial resolution $\Delta$.} }
\label{fig:statiocon}
\end{figure}

\begin{figure}
\includegraphics[width=\textwidth]{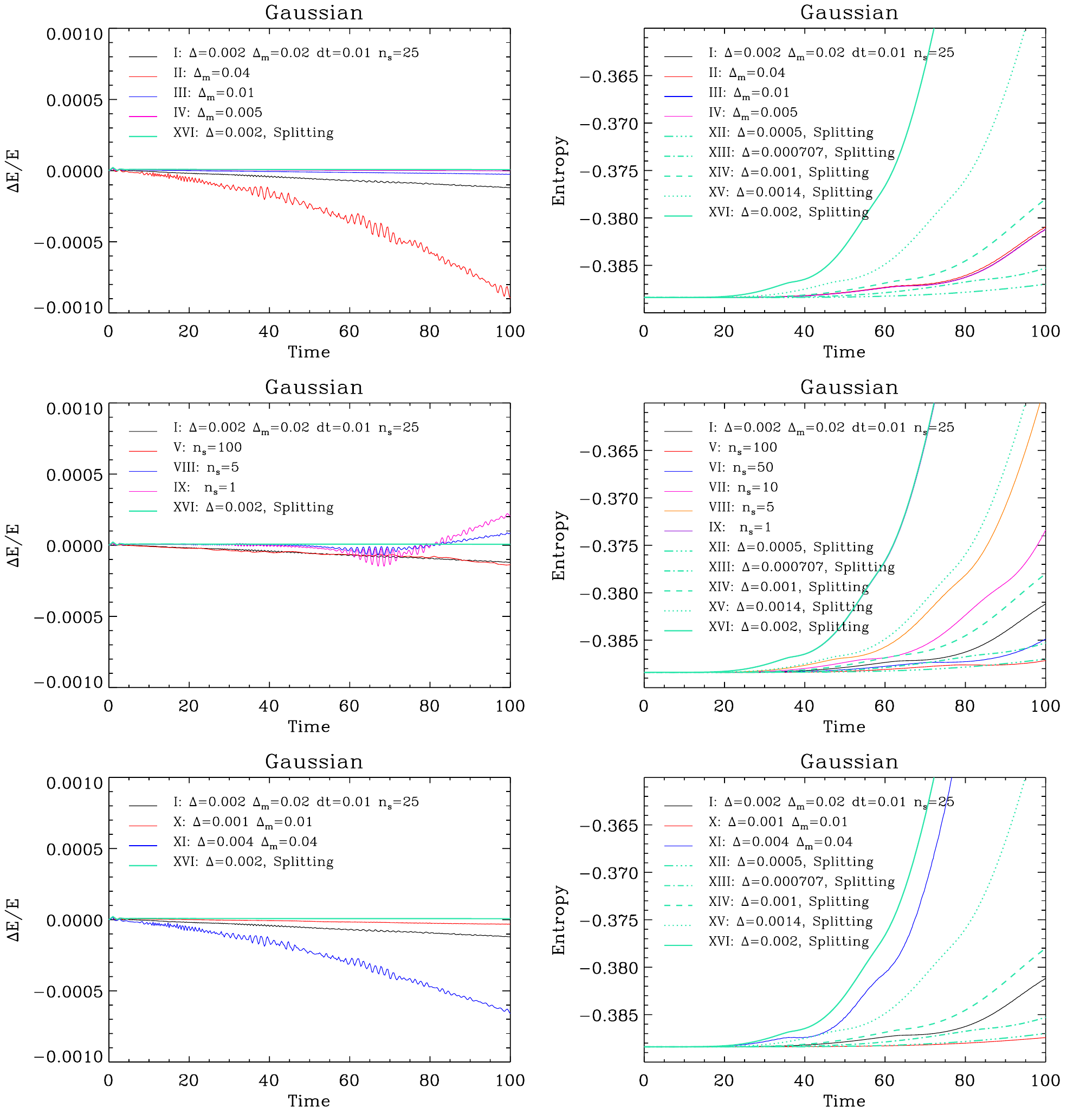}
\caption[]{Conservation of energy (left panels) and entropy (right panels) in the Gaussian runs, similarly as in Fig.~\ref{fig:statiocon}. { On left panels, energy conservation of the splitting algorithm is plotted only for run XVI,  but other runs  (XII, XIII, XIV and XV) would provide nearly indistinguishable results given the interval of values investigated. }}
\label{fig:gaussiancon}
\end{figure}

\begin{figure}
\includegraphics[width=\textwidth]{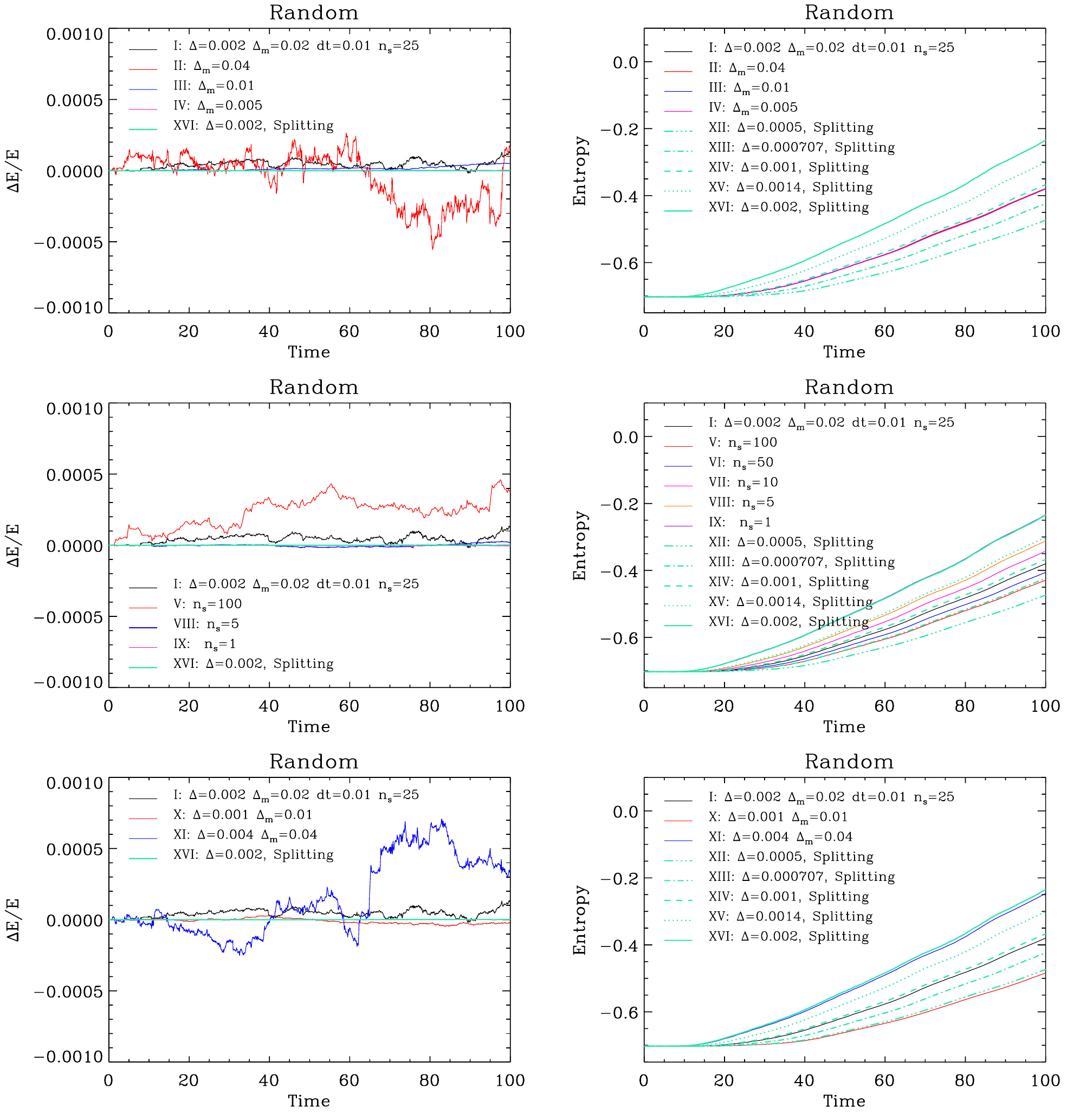}
\caption[]{Conservation of energy (left panels) and entropy (right panels) in the random set of halos runs, similarly as in Fig.~\ref{fig:statiocon}. { Again, energy conservation of the splitting algorithm is plotted only for run XVI, because other runs would provide nearly indistinguishable results. }}
\label{fig:randomcon}
\end{figure}

\begin{figure}
\includegraphics[width=\textwidth]{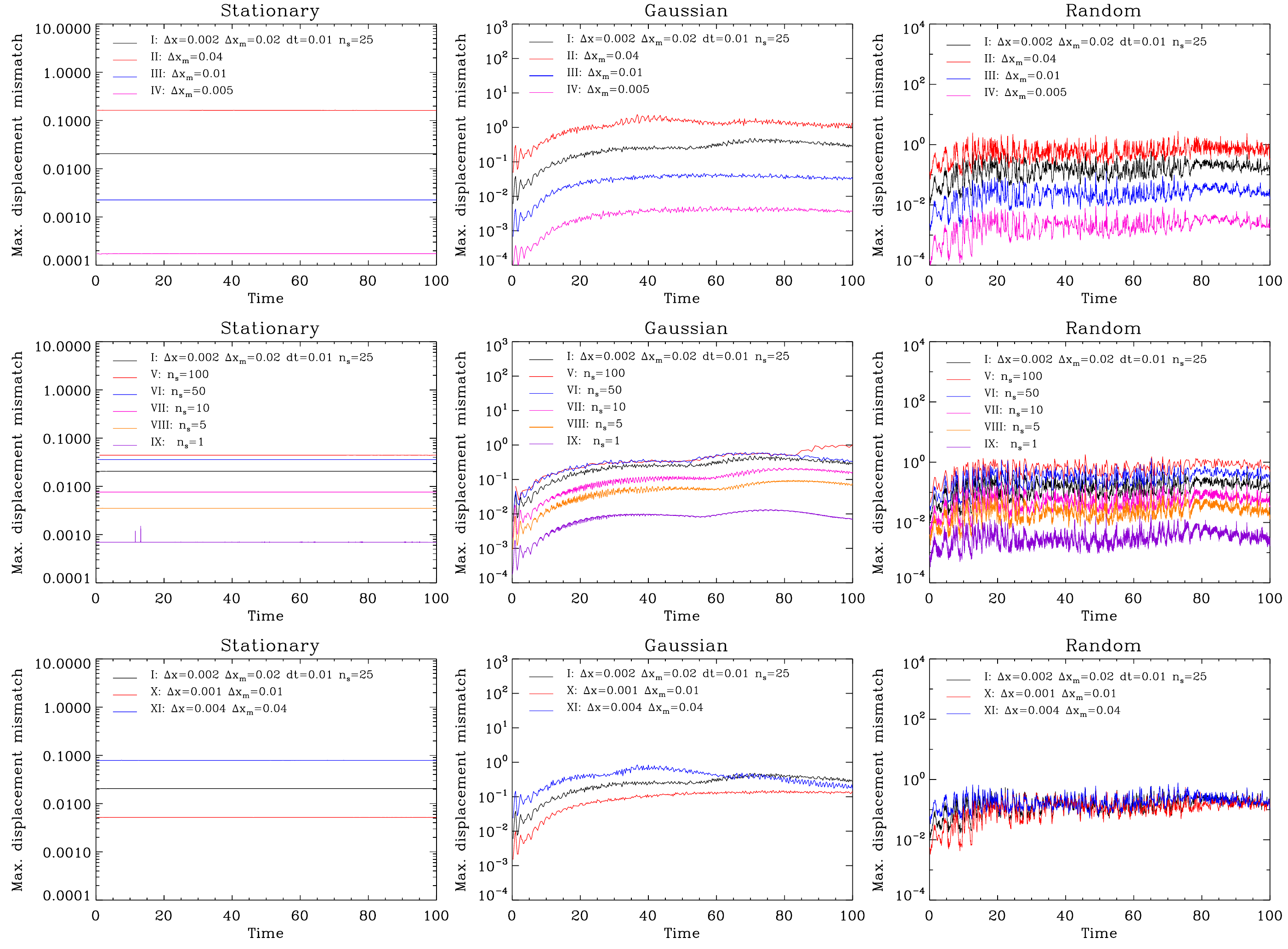}
\caption[]{{ The maximum reconstructed displacement mismatch between neighbouring elements of metric, $E_{\rm map}$, as defined in \S~\ref{sec:estierr}, as a function of time for the {\tt Vlamet} runs considered in the energy and entropy plots shown in previous figures. $E_{\rm map}$ is expressed in units of spatial resolution, $\Delta$. }}
\label{fig:errordis}
\end{figure}

Figures \ref{fig:statiocon}, \ref{fig:gaussiancon} and \ref{fig:randomcon} show $E$ and $S$ in the simulations we performed for the stationary, the Gaussian and the random set of halos, respectively.  These figures allow us to test thoroughly the performances of {\tt Vlamet} and to compare it to the splitting algorithm. They are supplemented with Figure~\ref{fig:errordis}, which displays the corresponding maximum mismatch between the Lagrangian positions predicates from neighbouring metric elements, as defined in \S~\ref{sec:estierr}.  


The most striking result when examining Figs.~\ref{fig:gaussiancon} and \ref{fig:randomcon} is the superior behaviour of {\tt Vlamet} compared to the standard splitting method with respect to entropy, in agreement with the preliminary conclusions driven from the analysis of Fig.~\ref{fig:gaussian_profile}. Indeed, to obtain comparable entropy conservation in {\tt Vlamet} and the splitting algorithm, it is required to have a higher spatial resolution in the latter. Of course, the results depend on the choice of $n_{\rm s}$ and of the ratio $\Delta x_{\rm m}/\Delta$. For instance, {\tt Vlamet} runs with $n_{\rm s}=25$ and $\Delta x_{\rm m}/\Delta=10$ perform as well as the splitting algorithm with twice the spatial resolution. But if $n_{\rm s}$ is increased to 100, it is required to increase the resolution by a factor as large as $2.8$ (middle right panel of Fig.~\ref{fig:randomcon}) or $4$ (middle right panel of Fig.~\ref{fig:gaussiancon}) in the splitting algorithm to perform as well as {\tt Vlamet}.  The price to pay might be energy conservation, which is in general not as good in {\tt Vlamet} as in the splitting algorithm, that preserves energy at a level better than $\sim 10^{-5}$ regardless of initial conditions and spatial resolution. However, energy conservation violation in {\tt Vlamet} can be kept under control with the appropriate combination of parameter $n_{\rm s}$ and ratio $\Delta_{\rm m}/\Delta$, as discussed further below. On the same ground, we notice, when examining Fig.~\ref{fig:statiocon}, that the splitting algorithm seems to preserve better the thermal equilibrium solution than {\tt Vlamet}, as already noticed on right panel of Fig.~\ref{fig:stationfxv}, unless $n_{\rm s}$ or/and $\Delta_{\rm m}/\Delta$ are chosen small enough. However, one has to be aware that the smoothness of the thermal equilibrium solution provides a very good ground for the splitting algorithm.

Further investigation of Figs.~\ref{fig:statiocon}, \ref{fig:gaussiancon}, \ref{fig:randomcon} and \ref{fig:errordis} allows us to study more in details how {\tt Vlamet} behaves with the various control parameters:

\begin{itemize}
\item Increasing overall resolution, i.e. decreasing $\Delta=\Delta x=\Delta y$ and $\Delta_{\rm m}=\Delta x_{\rm m}=\Delta y_{\rm m}$ by some factor improves both energy conservation and entropy conservation, as can be seen from last line of panels of all the figures, as expected. Note that adopting $\Delta x=\Delta y$ and $\Delta x_{\rm m}=\Delta y_{\rm m}$ as we did in this work is not necessarily optimal, but is a natural choice in the code units if total mass of the system is of the order of unity. 
\item  Changing the resampling frequency parameter $n_{\rm s}$ affects both entropy and energy, but to a much larger extent the former than the latter. In particular, increasing $n_{\rm s}$ up to 100 clearly improves entropy at an acceptable cost for energy conservation, of which the violation remains below the $\sim 5\times 10^{-4}$ level, as can be seen from the second line of panels of Figs.~\ref{fig:statiocon}, \ref{fig:gaussiancon} and \ref{fig:randomcon}.  This is not surprising: performing less frequent resamplings of the phase-space distribution function decreases diffusion and aliasing, which is good for entropy conservation, but it worsens energy conservation, because (i) the error on the reconstruction of Lagrangian coordinates increases (middle line of panels in Fig.~\ref{fig:errordis}) and (ii) the phase-space distribution function is sampled more approximately during subcycles. While effect (i) can be reduced by augmenting the metric elements density as discussed below, effect (ii) always remains, which prevents arbitrarily large values of $n_{\rm s}$. 
\item Sparse sampling the elements of metric, i.e. increasing $\Delta_{\rm m}$, augments the error in the Lagrangian coordinates reconstruction (first line of panels in Fig.~\ref{fig:errordis}). This affects energy conservation but not so much the entropy (except for the stationary simulations, which nevertheless preserve entropy at a level better than $10^{-3}$). Entropy is indeed mainly sensitive to spatial resolution, i.e. the value of $\Delta$, and as stated just above, to the frequency of resamplings. For the experiments we did, we find that the ratio $\Delta_{\rm m}/\Delta$ should not exceed a factor 10, and the smaller it is, the better the results, obviously. For instance, using $\Delta_{\rm m}/\Delta \la 5$ leads to energy conserved better than $\sim 5\times 10^{-5}$ for $\Delta=0.002$ and $n_{\rm s}=25$. Being aware that the gain in memory is already a factor $2.5^6=244$ in 6D when sparse sampling the elements of metric by a factor $2.5$ in each direction compared to the computational grid,  it is obviously wise to keep the elements of metric sampling sufficiently dense at the price of a small increase of the computational cost (a factor of the order of 1.4 in two-dimensional phase-space for $\Delta_{\rm m}/\Delta=2.5$, according to fifth column of Table~\ref{tab:tab}). Furthermore, a higher density of metric elements allows for a larger value of $n_{\rm s}$. 
\end{itemize}

\section{Conclusion}
\label{sec:conclusion}
 We have proposed and tested a new Vlasov solver, {\tt Vlamet}, which uses small elements of metric to carry information about the flow and its deformation. While slightly more costly in 2-dimensional phase-space than standard semi-Lagrangian solvers such as the splitting algorithm at the same resolution, this code is intrinsically less diffusive due to seldom actual resampling of the phase-space distribution function, as explicitly demonstrated by our numerical experiments. For the tests we performed, we indeed found that to have comparable level of details in the splitting algorithm and in {\tt Vlamet}, it was needed to increase the spatial resolution by a factor 2 to 4 in the splitting algorithm runs, and this factor might turn to become even larger after additional optimisation: some work is indeed still needed to understand how to use {\tt Vlamet} in the best way, while keeping errors, in particular deviations from energy conservation, under control. Indeed, despite the numerous runs we did, we did not span all the possible range of control parameters of {\tt Vlamet}. In particular, we did not explore thoroughly the case when the separation between elements of metric approaches the spatial resolution, e.g $2.5 \la \Delta_{\rm m}/\Delta \la 5$, which should allow for even more seldom resamplings than what we achieved. 

The gain in effective spatial resolution of {\tt Vlamet} compared to the splitting algorithm compensates largely for the cost of the code and makes our method potentially extremely competitive in 4 and 6 dimensional phase-space. Our current implementation treats isolated self-gravitating systems in 2 dimensional phase-space and uses parallel acceleration on shared memory architecture with the OpenMP library. The next step, quite relevant to galactic dynamics consists in generalising the code to 4 and 6 dimensional phase-space. This extension will require the introduction of adaptive mesh refinement with e.g. a k-d tree structure to focus on regions of phase-space of interest as well as parallel programming on distributed memory architecture. This should not present, in principle, any major difficulty, as discussed in \S~\ref{sec:perf}.

Note also that generalisation to plasmas is obvious and could represent a way to improve the accuracy of current codes. 

Still, our implementation remains quite naive and some more work is needed for full optimisation. In particular, we use simple finite difference method to compute gradient and Hessian of the force, which calls for improvements. Also, at the time of resampling, we employ the traditional third order spline interpolation which is clearly not free of defects, in particular aliasing effects that induce strong deviations from positivity of the phase-space distribution function (while, nevertheless not modifying the good dynamical behaviour of the system).  One possible way to improve over cubic B-splines could consist in employing a discontinuous Galerkin representation \citep[see, e.g.][]{Cockburn2000,Qiu2011,Besse2017}, where the phase-space distribution function is described in each computational cell by some polynomial of some order. Continuity is not necessarily enforced between neighbouring cells, which provides more flexibility. However, while the splitting algorithm allows one to decompose the dynamical set up in individual one dimensional problems, this is not possible in {\tt Vlamet}. We have to find a way, at the moments of remapping, to reconstruct a new discontinuous Galerkin representation by composing these local polynomials defined in a four- or six-dimensional space with the non linear displacement field provided by the metric elements, which remains a challenging task.

\section*{Acknowledgements}
SC greatly acknowledges hospitality of Yukawa Institute for Theoretical Physics (YITP), where part of this work has been carried out. We thank N. Besse for fruitful discussions. We also acknowledge the support of YITP in organising the workshop ``Vlasov-Poisson: towards numerical methods without particles'' in Kyoto, funded by grant YITP-T-15-02, ANR grant ANR-13-MONU-0003 and by Institut Lagrange de Paris (ANR-10-LABX-63 and ANR-11-IDEX-0004-02). This work was supported in part by ANR grant ANR-13-MONU-0003.
\bibliographystyle{jpp}

\end{document}